\documentclass[preprint2]{emulateapj} 
\usepackage{amsmath}
\usepackage{natbib}
\usepackage{graphicx}
\usepackage{epsf}
\usepackage{color}
\input{epsf}
\usepackage{epsfig}
\DeclareGraphicsExtensions{.jpg,.pdf,.png,.eps,.ps}
\graphicspath{{FIGURES/}}

\newcommand{\hmsun}{h^{-1} \; M_{\odot}}

\newcommand{\planck}{{\it Planck}/HFI}
\newcommand{\herschel}{{\it Herschel}}

\newcommand{\sqdeg}{\mbox{deg$^2$}}
\newcommand{\chisq}{\ensuremath{\chi^2}}
\newcommand{\delchisq}{\ensuremath{\Delta\chi^2}}
\newcommand{\zcenter}{\ensuremath{z_{\rm c}}}

\newcommand{\ltsima}{$\; \buildrel < \over \sim \;$}
\newcommand{\ltsim}{\lower.5ex\hbox{\ltsima}}

\newcommand{\amplitudeletter}{D_{3000}^}
\newcommand{\shapeletter}{D_{\ell}^}
\newcommand{\muksq}{$\mu {\rm K}^2$}
\newcommand{\atsz}{A_{\rm tSZ}}
\newcommand{\aksz}{A_{\rm kSZ}}

\def\n{{\bf  \hat n}}
\def\q{{\bf q}}

\newcommand{\degsq}{deg$^2$}

\begin{document}

\title{A measurement of secondary cosmic microwave background anisotropies with two years of South Pole Telescope observations}

\author{
 C.~L.~Reichardt,\altaffilmark{1}
  L.~Shaw,\altaffilmark{2} 
   O.~Zahn,\altaffilmark{3} 
 K.~A.~Aird,\altaffilmark{4}
 B.~A.~Benson,\altaffilmark{5,6}
 L.~E.~Bleem,\altaffilmark{5,7}
 J.~E.~Carlstrom,\altaffilmark{5,6,7,8,9}
 C.~L.~Chang,\altaffilmark{5,6,9}
 H.~M. Cho,\altaffilmark{10} 
 T.~M.~Crawford,\altaffilmark{5,8}
 A.~T.~Crites,\altaffilmark{5,8}
 T.~de~Haan,\altaffilmark{11}
 M.~A.~Dobbs,\altaffilmark{11}
 J.~Dudley,\altaffilmark{11}
 E.~M.~George,\altaffilmark{1}
 N.~W.~Halverson,\altaffilmark{12}
 G.~P.~Holder,\altaffilmark{11}
 W.~L.~Holzapfel,\altaffilmark{1}
 S.~Hoover,\altaffilmark{5,7}
 Z.~Hou,\altaffilmark{13}
 J.~D.~Hrubes,\altaffilmark{4}
 M.~Joy,\altaffilmark{14}
  R.~Keisler,\altaffilmark{5,7}
 L.~Knox,\altaffilmark{13}
 A.~T.~Lee,\altaffilmark{1,15}
 E.~M.~Leitch,\altaffilmark{5,8}
 M.~Lueker,\altaffilmark{16}
 D.~Luong-Van,\altaffilmark{4}
 J.~J.~McMahon,\altaffilmark{17}
 J.~Mehl,\altaffilmark{5}
 S.~S.~Meyer,\altaffilmark{5,6,7,8}
 M.~Millea,\altaffilmark{13}
 J.~J.~Mohr,\altaffilmark{18,19,20}
 T.~E.~Montroy,\altaffilmark{21}
 T.~Natoli,\altaffilmark{5,7}
 S.~Padin,\altaffilmark{5,8,16}
 T.~Plagge,\altaffilmark{5,8}
 C.~Pryke,\altaffilmark{5,7,8,22}
 J.~E.~Ruhl,\altaffilmark{21}
 K.~K.~Schaffer,\altaffilmark{5,6,23}
 E.~Shirokoff,\altaffilmark{1} 
 H.~G.~Spieler,\altaffilmark{15}
 Z.~Staniszewski,\altaffilmark{21}
 A.~A.~Stark,\altaffilmark{24}
 K.~Story,\altaffilmark{5,7}
 A.~van~Engelen,\altaffilmark{11}
 K.~Vanderlinde,\altaffilmark{11}
 J.~D.~Vieira,\altaffilmark{16} and 
 R.~Williamson\altaffilmark{5,8} 
}

\altaffiltext{1}{Department of Physics,
University of California, Berkeley, CA, USA 94720}
\altaffiltext{2}{Department of Physics, Yale University, P.O. Box 208210, New Haven,
CT, USA 06520-8120}
\altaffiltext{3}{Berkeley Center for Cosmological Physics,
Department of Physics, University of California, and Lawrence Berkeley
National Labs, Berkeley, CA, USA 94720}
\altaffiltext{4}{University of Chicago,
5640 South Ellis Avenue, Chicago, IL, USA 60637}
\altaffiltext{5}{Kavli Institute for Cosmological Physics,
University of Chicago, 5640 South Ellis Avenue, Chicago, IL, USA 60637}
\altaffiltext{6}{Enrico Fermi Institute,
University of Chicago,
5640 South Ellis Avenue, Chicago, IL, USA 60637}
\altaffiltext{7}{Department of Physics,
University of Chicago,
5640 South Ellis Avenue, Chicago, IL, USA 60637}
\altaffiltext{8}{Department of Astronomy and Astrophysics,
University of Chicago,
5640 South Ellis Avenue, Chicago, IL, USA 60637}
\altaffiltext{9}{Argonne National Laboratory, 9700 S. Cass Avenue, Argonne, IL, USA 60439}
\altaffiltext{10}{NIST Quantum Devices Group, 325 Broadway Mailcode 817.03, Boulder, CO, USA 80305}
\altaffiltext{11}{Department of Physics,
McGill University, 3600 Rue University, 
Montreal, Quebec H3A 2T8, Canada}
\altaffiltext{12}{Department of Astrophysical and Planetary Sciences and Department of Physics,
University of Colorado,
Boulder, CO, USA 80309}
\altaffiltext{13}{Department of Physics, 
University of California, One Shields Avenue, Davis, CA, USA 95616}
\altaffiltext{14}{Department of Space Science, VP62,
NASA Marshall Space Flight Center,
Huntsville, AL, USA 35812}
\altaffiltext{15}{Physics Division,
Lawrence Berkeley National Laboratory,
Berkeley, CA, USA 94720}
\altaffiltext{16}{California Institute of Technology, MS 249-17, 1216 E. California Blvd., Pasadena, CA, USA 91125}
\altaffiltext{17}{Department of Physics, University of Michigan, 450 Church Street, Ann  
Arbor, MI, USA 48109}
\altaffiltext{18}{Department of Physics,
Ludwig-Maximilians-Universit\"{a}t,
Scheinerstr.\ 1, 81679 M\"{u}nchen, Germany}
\altaffiltext{19}{Excellence Cluster Universe,
Boltzmannstr.\ 2, 85748 Garching, Germany}
\altaffiltext{20}{Max-Planck-Institut f\"{u}r extraterrestrische Physik,
Giessenbachstr.\ 85748 Garching, Germany}
\altaffiltext{21}{Physics Department, Center for Education and Research in Cosmology 
and Astrophysics, 
Case Western Reserve University,
Cleveland, OH, USA 44106}
\altaffiltext{22}{Department of Physics, University of Minnesota, 116 Church Street S.E. Minneapolis, MN, USA 55455}
\altaffiltext{23}{Liberal Arts Department, 
School of the Art Institute of Chicago, 
112 S Michigan Ave, Chicago, IL, USA 60603}
\altaffiltext{24}{Harvard-Smithsonian Center for Astrophysics,
60 Garden Street, Cambridge, MA, USA 02138}

\email{cr@bolo.berkeley.edu}
 
\begin{abstract}
We present the first three-frequency South Pole Telescope (SPT) cosmic microwave background (CMB) power spectra. 
The band powers presented here cover angular scales $2000 < \ell < 9400$ in frequency bands centered at 95, 150, and 220\,GHz. 
At these frequencies and angular scales, a combination of the primary CMB anisotropy, thermal and kinetic Sunyaev-Zel'dovich (SZ) effects, radio galaxies, and cosmic infrared background (CIB) contributes to the signal. 
We combine \planck\ and SPT data at 220\,GHz to constrain the amplitude and shape of the CIB power spectrum and find strong evidence for non-linear clustering. 
We explore the SZ results using a variety of cosmological models for the 
CMB and CIB anisotropies and find them to be robust with one exception:
allowing for spatial correlations between the thermal SZ effect and CIB significantly degrades the SZ constraints. 
Neglecting this potential correlation, we find the thermal SZ power at 150\,GHz and $\ell = 3000$ to be $3.65 \pm 0.69 \,\mu{\rm K}^2$, and set an upper limit on the kinetic SZ power to be less than $2.8 \,\mu{\rm K}^2$ at 95\% confidence. 
When a correlation between the thermal SZ and CIB is allowed, we constrain a linear combination of thermal and kinetic 
SZ power: 
$D_{3000}^{\rm tSZ} + 0.5 D_{3000}^{\rm kSZ} = 4.60 \pm 0.63 \,\mu{\rm K}^2$, consistent with earlier measurements. 
We use the measured thermal SZ power and an analytic, thermal SZ model calibrated with simulations to determine $\sigma_8 = 0.807 \pm 0.016$. 
Modeling uncertainties involving the astrophysics of the intracluster medium rather than the statistical uncertainty in the measured band powers are the dominant source of uncertainty on $\sigma_8$. 
We also place an upper limit on the kinetic SZ power produced by patchy reionization; a companion paper uses these limits to constrain the reionization history of the Universe.

\end{abstract}

\keywords{cosmology -- cosmology:cosmic microwave background -- cosmology:diffuse radiation--  cosmology: observations -- large-scale structure of universe }

\bigskip\bigskip

\section{Introduction}

The current generation of ground-based cosmic microwave background
(CMB) experiments enables researchers to probe, for the first
time, the power spectra of the thermal and kinetic Sunyaev-Zel'dovich
(SZ) effects \citep{sunyaev72,sunyaev80b}.  The South Pole Telescope
\citep[SPT,][]{carlstrom11} and the Atacama Cosmology Telescope
\citep[ACT,][]{fowler07} have sufficient sensitivity, angular
resolution, and frequency coverage to measure the power spectrum of the thermal SZ (tSZ)
effect and to place increasingly strong upper limits on contributions from the kinetic SZ
(kSZ) effect \citep{lueker10, shirokoff11, das11b}.  These secondary
temperature anisotropies in the CMB arise when CMB photons
are scattered by the inhomogeneous distribution of free electrons well after
recombination.  As such, they depend sensitively on both the growth of
structure and the details of reionization.

The tSZ effect occurs when CMB photons inverse Compton scatter off hot electrons, predominantly those  gravitationally bound in the potential wells of galaxy clusters.  
The scattering tends to increase the photon energy, distorting the CMB black-body spectrum with a loss of photons below approximately 217~GHz and an excess at higher frequencies.  
The kSZ effect occurs when CMB photons are Doppler shifted by the bulk velocity of electrons in intervening ionized gas.  
The kSZ effect changes the apparent CMB temperature while maintaining the black-body spectrum.  
We can discriminate between the two effects based on their frequency dependence.

The anisotropy power due to the tSZ effect is highly 
sensitive to the normalization of the matter power spectrum, commonly
parametrized by the RMS of the $z=0$ mass distribution on 8 Mpc/$h$ scales,
$\sigma_8$ \citep{komatsu02}.  As a measure of structure growth, the tSZ power spectrum
can provide independent constraints on cosmological parameters and
improve the precision with which they are determined.  For instance, a
measurement of the tSZ power spectrum can be used to determine the sum
of the neutrino masses by breaking the degeneracy between $\sigma_8$
and neutrino mass that exists with the CMB data alone.  The tSZ power
spectrum can also be used to constrain non-standard cosmological
models, for example, placing limits on the range of allowed early dark
energy models \citep{alam11}.

The tSZ power spectrum is challenging to model accurately because it
includes significant contributions from galaxy clusters spanning a
wide mass and redshift range.  Modeling uncertainties arise from both
non-gravitational heating effects in low-mass clusters and the limited
body of observational data on high-redshift clusters.  Recent models that
vary in their treatment of cluster gas physics differ in amplitude by
up to 50\% for a given cosmology \citep{shaw10, trac11, battaglia10}.

The post-reionization kSZ power spectrum is expected to be easier to
model than the tSZ spectrum.  As it is not weighted by the gas
temperature, the kSZ power spectrum depends less on 
non-linear effects that complicate models of the tSZ signal. 
The total kSZ power spectrum does, however, depend strongly on the details of
reionization, which are not yet well understood.  In the standard
picture of reionization, ionized bubbles form around the first stars
or quasars. These bubbles eventually overlap and merge, leading to a
fully ionized universe. 
The ionized bubbles impart a Doppler shift on scattered CMB photons proportional to their line-of-sight velocity.  
This so-called
``patchy'' reionization kSZ signal depends on the duration of
reionization as well as the distribution of bubble sizes \citep{gruzinov98,knox98}.
Detection of the kSZ power spectrum or even an upper limit on its
amplitude would lead to interesting constraints on the epoch of
reionization \citep{zahn05}.

The first detection of the combined  tSZ and kSZ spectra was presented by \citet[hereafter L10]{lueker10} using data from the SPT. 
Since then, improved measurements of the power spectra have been reported by \citet[hereafter S11]{shirokoff11} based on 150 and 220\,GHz data for $200\,{\rm deg}^2$  of sky. 
S11 report a $4.5\,\sigma$ detection of SZ power at 152\,GHz at a multipole of $\ell=3000$ with an amplitude  of $D^{\rm tSZ}_{3000} + 0.5 D^{\rm kSZ}_{3000} = 4.5\pm1.0\,\mu{\rm K}^2$.\footnote{Throughout this work, the unit $\textrm{K}$ refers to equivalent fluctuations in the CMB temperature, i.e.,~the temperature fluctuation of a 2.73$\,$K blackbody that would be required to produce the same power fluctuation.  The flux conversion factor is given by the derivative of the blackbody spectrum, $\frac{dB}{dT}$, evaluated at 2.73$\,$K.}  
The 95\% confidence upper limits on each component were $D^{\rm tSZ}_{3000} < 5.3\,\mu{\rm K}^2$ and $D^{\rm kSZ}_{3000} < 6.5\,\mu{\rm K}^2$. 
The measured tSZ power is less than predicted by some models, suggesting that either these models over-predict the tSZ power or $\sigma_8$ is significantly less than the WMAP preferred value of $\sim$\,0.8 \citep{larson11}.

The ACT collaboration has also measured the high-$\ell$ power spectrum \citep{das11b}.
\citet{dunkley11} use the \citet{das11b} bandpowers  to measure  the sum of the tSZ and kSZ power at $\ell=3000$ and $148\,$GHz to be $6.8 \pm 2.9 \,\mu {\rm K}^2$. 
The SPT and ACT bandpowers and resulting constraints on SZ power are consistent within the reported
uncertainties.

The sky power at millimeter
wavelengths and arcminute scales is dominated by 
the signal from bright synchrotron sources, with additional contributions from nearby infrared galaxies and a population of high-redshift, gravitationally lensed,
dusty, star-forming galaxies (DSFGs)  \citep{vieira10}. 
After these identifiable sources are masked, the power spectrum is dominated by the cosmic
infrared background (CIB), the bulk of which is contributed by 
unlensed DSFGs below the detection threshold of instruments like SPT and ACT \citep{lagache05}.
These DSFGs are important both as a foreground for CMB experiments and as a tool to understand the history of star and galaxy formation \citep[e.g.][]{bond86,bond91b,knox01,lagache05}.  
The power from DSFGs has both Poisson and clustered components, with distinct angular scale dependencies.  
Measurements of CIB power at multiple millimeter wavelengths can be combined to constrain the spectral index of the CIB anisotropy. 
Note that this depends on both the spectral energy distribution of individual DSFGs and their redshift distribution. 
S11 constrained the spectral index of the DSFGs between 150 and 220\,GHz to be $\alpha = 3.58 \pm 0.09$.  
\citet{dunkley11} also detected significant power attributed to clustered DSFGs in the ACT data and found a preferred DSFG spectral index of $3.69 \pm 0.14$, consistent with the S11 results. 
The \citet{planck11-6.6_arxiv} combined the \planck\ CIB bandpowers with the S11 bandpowers to show that the combined dataset is inconsistent with only the Poisson and linear clustering terms -- non-linear clustering is required.

This is the fourth SPT power spectrum analysis focused on CIB and secondary CMB anisotropies.  
Here, we improve upon the previous work by L10, \citet{hall10} and S11 in two ways. 
First, we are analyzing three-frequency data (with six cross-spectra) where earlier analyses used two frequency data with three cross-spectra. 
The additional frequency information allows the separation of CIB and SZ components. 
Second, we are analyzing four times more sky area ($800\,$deg$^2$) with the concomitant reduction in bandpower uncertainties. 
A companion paper, \citet[Z11]{zahn11b}, interprets the derived kSZ constraints in light of the epoch of reionization.

The paper is organized as follows. 
In \S\ref{sec:dataanalysis}, we present the analysis and the data to which it is applied. 
We examine the results of tests for systematic errors in the data in \S\ref{sec:jackknife}. 
The resulting power spectrum is presented in \S\ref{sec:bandpowers}. 
The cosmological modeling is discussed in \S\ref{sec:model} and basic results in \S\ref{sec:results}. 
We explore constraints on the tSZ effect in \S\ref{sec:tszconstraint}, on the kSZ effect in \S\ref{sec:kszconstraint}, and on the CIB in \S\ref{sec:cibconstraint}. 
We conclude in \S\ref{sec:conclusion}.

\section{Data and analysis}
\label{sec:dataanalysis}

We present bandpowers from 800 \degsq\ of sky observed with the SPT at 95, 150, and 220\,GHz. 
Bandpowers are estimated using a pseudo-${\it C}_\ell$ cross-spectrum method. 
We calibrate the data by comparing the power spectrum to seven-year WMAP (WMAP7, \citealt{larson11}) bandpowers.

\subsection{Data}
\label{subsec:data}

The 800 \degsq\ of sky analyzed in this work were observed by the SPT during the 2008 and 2009 Austral winters. 
The SPT is an off-axis Gregorian telescope with a 10 meter diameter
primary mirror located at the South Pole.  
The receiver is equipped with 960 horn-coupled spiderweb bolometers with 
superconducting transition edge sensors.
The telescope and receiver are discussed in more detail in \citet{ruhl04}, \citet{padin08}, and \citet{carlstrom11}. 

Each field was observed to an approximate depth of 18 $\mu$K-arcmin at 150\,GHz. 
The noise levels at 95 and 220\,GHz vary significantly between the two years as the focal plane was refurbished to maximize sensitivity to the tSZ effect before the 2009 observing season. 
This refurbishment added science-quality 95\,GHz detectors and removed half the 220\,GHz detectors. 
As a result, the 2009 data are comparatively shallower at 220\,GHz, but much deeper at 95\,GHz. 
The 2009 sky coverage accounts for approximately 75\% of the total sky used in this analysis. 
When calculating the power spectra, the 2009 data has a statistical weight of 100\% at 95\,GHz, 73\% at 150\,GHz and 43\% at 220\,GHz.

\subsection{Beams and Calibration}
\label{sec:beamcal}

The SPT beams are measured for each year using a combination of bright point sources in each field, Venus, and Jupiter as described in S11. 
The main lobes of the SPT beam at 95, 150, and 220$\,$GHz are well-represented by 1.7$^\prime$, 1.2$^\prime$, and 1.0$^\prime$ FWHM Gaussians.
As in S11, we calculate the principal components of the full covariance matrix of this measurement and find that three eigenvectors are adequate to represent $>$99\% of the covariance.
We treat the eigenvectors from different years as independent. 
We have tested the opposite extreme, treating them as completely correlated, and have found minimal impact on the science results. 
The beam uncertainty is included in the bandpower covariance matrix as described in \S\ref{subsec:beamunc}.

The observation-to-observation relative calibrations of the time-ordered data (TOD) are determined from repeated measurements of a galactic H{\small \,II} region, RCW38. 
As in \citet[K11]{keisler11}, the absolute calibration at each frequency is determined by comparing the SPT and WMAP7 bandpowers over the multipole range $\ell \in [650,1000]$. 
We use the same $\ell$-bins for each experiment: seven bins with $\Delta\ell = 50$. 
For the absolute calibration, we do not weight individual
modes within each bin (as we will do in section \ref{sec:bandpowerest}). 
A uniform weighting is used because these large angular scales are dominated by
sample variance rather than instrumental and atmospheric noise. 
This calibration method is model-independent, requiring only that the CMB power in the SPT fields is statistically representative of the all-sky power. 
We estimate the uncertainty in the SPT power calibration at 95, 150, and 220\,GHz to be 3.5\%, 3.2\%, and 4.8\% respectively. 
These uncertainties are highly correlated because the main sources of error, WMAP7 bandpower errors and SPT sample variance, are nearly identical between bands.

\subsection{Bandpower estimation}
\label{sec:bandpowerest} 

We use a pseudo-$C_\ell$ method to estimate the bandpowers \citep{hivon02}. 
 In pseudo-$C_\ell$ methods, bandpowers are estimated directly from the spherical harmonic transform of the map after correcting for effects such as TOD filtering, beams, and finite sky coverage. 
We process the data using a cross spectrum based analysis \citep{polenta05, tristram05} to eliminate noise bias.  
Our scan strategy produces of order 100 complete but noisy maps of each field, and so is ideally suited to a cross-spectrum analysis. 
We use a flat-sky approximation. 
Before Fourier transforming, each map is apodized by a window designed to mask point sources detected at greater than $>5\,\sigma$ (6.4\,mJy) at 150\,GHz, and to avoid sharp edges at the map borders. 
Beam and filtering effects are corrected for according to the formalism in the MASTER algorithm \citep{hivon02}.  
We report the bandpowers in terms of $\mathcal{D}_\ell$, where
\begin{equation}
\mathcal{D}_\ell=\frac{\ell\left(\ell+1\right)}{2\pi} C_\ell\;.
\end{equation}

Details of the bandpower estimator have been presented in previous SPT papers (L10, S11, K11). In the following sections, we will only note differences between the current analysis and what was done by S11.

\subsubsection{Simulations}

Simulations are used to estimate the transfer function and sample variance. 
For each field, we construct simulated observations of 100 sky realizations that have been smoothed by the appropriate beam. 
The sky realizations are sampled using the SPT pointing information, filtered identically to the real data, and processed into maps.

Each simulated sky is a Gaussian realization of the sum of the  best-fit lensed WMAP7 $\Lambda$CDM primary CMB model, a kSZ model, and point source contributions. 
Note that this approach neglects non-Gaussianity in the kSZ and radio source contributions and may slightly underestimate the sample variance as a result. 
\citet{millea11} argue this is negligible for SPT  since the non-Gaussian signals are small and, on the relevant angular scales, the bandpower uncertainties are dominated by instrumental noise. 
The kSZ power spectrum is taken from the \citet{sehgal10} simulations and has an amplitude of $2.05\, \mu{\rm K}^2$ at $\ell=3000$. 
We include both Poisson and clustered point sources. 
The Poisson contribution reflects both radio source and DSFG populations.
 The amplitude of the radio source term is set by the \citet{dezotti05} model source counts to an amplitude $D_{3000}^{r} = 1.28\, \mu{\rm K}^2$ at $150\,$GHz with an assumed spectral index of $\alpha_r=-0.6$. 
 The amplitude of the Poisson DSFG term at 150\,GHz is $D_{3000}^{p} = 7.7 \,\mu{\rm K}^2$. 
 Finally, the clustered DSFG component is modeled by a $D_\ell \propto \ell$ term normalized to $D_{3000}^{p} = 5.9 \,\mu{\rm K}^2$ at 150\,GHz. 
 The DSFG terms have an assumed spectral index of 3.6.
The amplitude of each component was selected to be consistent with the S11 bandpowers. 

\subsubsection{Covariance estimation and conditioning}

The bandpower covariance matrix includes two terms: sample variance and instrumental noise variance.  
The sample variance is estimated with the signal-only Monte Carlo bandpowers. 
The instrumental noise variance is calculated from the distribution of the cross-spectrum bandpowers $D^{\nu_i\times\nu_j,AB}_b$ between observations A and B, and frequencies $\nu_i$ and $\nu_j$, as described in L10. 
We expect some statistical uncertainty of the form  
\begin{equation}
\left<\left(\textbf{C}_{bb^\prime}-\left<\textbf{C}_{bb^\prime}\right>\right)^2\right>=\frac{\textbf{C}_{bb^\prime}^2+\textbf{C}_{bb}\textbf{C}_{b^\prime b^\prime}}{n_{obs}}
\end{equation}
in the estimated bandpower covariance matrix. 
Here, $n_{obs}$ is the number of observations that go into the covariance estimate. 
This uncertainty on the covariance is expected to be significantly higher than the true covariance for almost all off-diagonal terms due to the dependence of the uncertainty on the (large) diagonal covariances.
We reduce the impact of this uncertainty by ``conditioning'' the covariance matrix. 

How we condition the covariance matrix is determined by the form we expect it to assume.
The covariance matrix can be viewed as a set of thirty-six square blocks, with the six on-diagonal blocks corresponding to the covariances of $95\times95$, $95\times150$, ... $220\times220\,$GHz spectra.
The bandpowers reported in Table \ref{tab:bandpowers} are obtained by first computing power spectra and covariance matrices for bins of width $\Delta\ell=200$ with a total of 37 initial $\ell$-bins, thus each of these blocks is an $37 \times 37$ matrix. 
The shape of the correlation matrix in each of these blocks is expected to be the same, as it is set by the apodization window.   
As a first step to conditioning the covariance matrix, we calculate the correlation matrices for the six on-diagonal blocks and average all off-diagonal elements at a fixed separation from the diagonal in each block,
\begin{equation}
\label{eqn:covcond}
\textbf{c}^\prime_{kk^\prime}=\frac{
\sum_{k_1-k_2=k-k^\prime} 
\frac{\widehat{\textbf{c}}_{k_1k_2}}{\sqrt{\widehat{\textbf{c}}_{k_1k_1}\widehat{\textbf{c}}_{k_2k_2}}}
}{\sum_{k_1-k_2=k-k^\prime} 1}.
\end{equation}
We set the correlation to zero at distances $\ell > 400$ where we find no evidence for correlations. 
We have tested that varying this condition by a factor of two in either direction does not affect the cosmological constraints. 
This averaged correlation matrix is then applied to all the blocks. 

The covariance matrix includes an estimate of the signal and, if both bandpowers share a common map, the noise variance. 
To illustrate the latter condition, instrumental noise is included in the blocks corresponding to ($150\times 150, 150\times 150$) and ($95\times 150, 150\times 220$) covariances, but not  the covariance between the $150 \times 150$ and $220\times220$ bandpowers. 
Given the map filtering, we neither expect nor observe correlated noise between different frequencies in the signal band. 

We impose one final condition on the estimated covariances of off-diagonal blocks for which the instrumental noise term is included, e.g.~the ($95\times 150, 150\times 220$) covariance. 
In these blocks, the noise on the covariance estimate for an $\ell$-bin can be large compared to the true covariance on small angular scales. 
As a result, we fit an approximate template based on the measured beam shapes from $\ell = 2000 - 4400$ and extrapolate this template to estimate the diagonal elements of the covariance matrix at $\ell > 4400$. 
We have tested zeroing these elements of the covariance matrix instead and found it does not affect the cosmological constraints. 

\subsubsection{Field weighting}

We construct a diagonal weight matrix for each field $i$, $w^{i}_{bb} = 1/C^i_{bb}$. 
We smooth the weights with $\Delta \ell = 1000$ at $\ell < 3000$ and $\Delta\ell = 2600$ on smaller scales to deal with uncertainty in the covariance estimate since we expect the optimal weights to vary slowly with angular scale.

The combined bandpowers are calculated from the individual-field bandpowers according to:
\begin{equation}
D_b = \sum_{i}D_{b}^{i}w^{i}_{bb}
\end{equation}
where 
\begin{equation}
{\sum_{i}w^{i}_{b b}} = 1.
\end{equation}

The covariance matrix is  combined likewise according to:
\begin{equation}
\textbf{C}_{bb^\prime} = \sum_{i}w^{i}_{bb}\textbf{C}_{bb^\prime}^{i}w^{i}_{b^\prime b^\prime}
\end{equation}

\subsubsection{Beam uncertainties}
\label{subsec:beamunc}

The procedure for estimating the beams and beam uncertainties is described in \S\ref{sec:beamcal}; here we discuss how the beam uncertainties are included in the cosmological analysis. 
In S11, three parameters per frequency band describing the beam uncertainties were marginalized over in the Monte Carlo Markov chain (MCMC) with a Gaussian prior. 
In this work, we instead follow the treatment laid out in K11 and include the beam uncertainties in the bandpower covariance matrix. 

Following K11, we estimate the ``beam correlation" matrix for each year:
\begin{equation}
\pmb{\rho}^{\rm beam}_{bb^\prime} = \left(\frac{\delta D_b}{D_b}\right) \left(\frac{\delta D_{b^\prime}}{D_{b^\prime}}\right)
\end{equation}
where 
\begin{equation}
\frac{\delta D_b}{D_b} = 1-\left(1+\frac{\delta B_b}{B_b}\right)^{-2},
\end{equation}
and $B_b$ is the Fourier transform of the beam map averaged across bin $b$. 

We combine the correlation matrix for each year according to the yearly weights. 
The correlation matrix is translated into a covariance matrix by multiplying by the measured bandpowers:

\begin{equation}
\textbf{C}^{\rm beam}_{bb^\prime} = \pmb{\rho}^{\rm beam}_{bb^\prime}D_{b}D_{b^\prime}.
\end{equation}

This beam covariance is added to the bandpower covariance matrix due to sample variance and instrumental noise.

\section{Jackknife Tests}
\label{sec:jackknife}

We test for systematic errors using a suite of ``jackknife" tests. 
A jackknife test consists of dividing the data into two halves and differencing them to remove true astrophysical signals.
The halves are chosen to maximize the signal  in the differenced spectrum due to a potential systematic bias. 
We then compare the differenced spectrum to the (nearly) zero expectation in the absence of a systematic error. 
To implement jackknife tests in the cross-spectrum framework, we calculate the cross-spectrum of differenced pairs of observations. 
More details on the implementation can be found in L10 and S11.
We look for systematic effects in the following dimensions.

\begin{itemize}

\item Scan direction: We difference the data based on whether the telescope is scanning to the left or right. 
This test is sensitive to direction- or turnaround-dependent effects, such as  detector time constants, or microphonics induced by telescope acceleration.

\item Azimuthal Range: We difference the data based on the azimuth at which it was observed. 
This is primarily a test for ground pickup. 
As discussed in S11, we select the azimuth ranges for the difference set based on the magnitude of observed ground emission on several degree scales. 

\item Time: We have two time-dependent tests. 
First, we compare the first and second halves of the data for each field. 
Second, we difference the bandpowers for 2008 from those of 2009. 
Unlike the other jackknife tests, this year-difference test is applied only to 150 and 220\,GHz; there are no 95\,GHz data from 2008. 
This year-difference is also implemented differently than the other jackknife tests. 
We calculate power spectra for each year and difference the bandpowers. 
Notably, this means that the test is less sensitive since the error bars must include sample variance and beam uncertainties. 
These two tests are sensitive to long-timescale drifts in the calibration, pointing, detector properties, or any other time-varying effect. 

\end{itemize}

\begin{figure}[t]\centering
\includegraphics[width=0.45\textwidth]{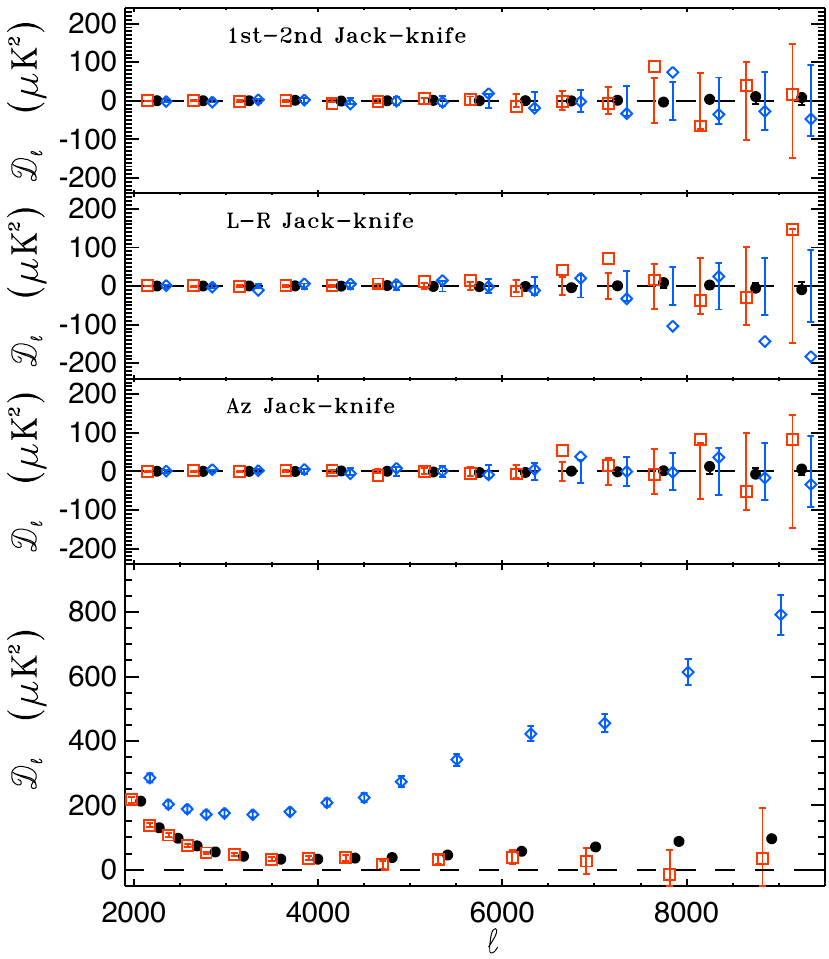}
  \caption[]{ The results of the jackknife tests applied to the SPT data. 
  {\bf Red squares} denote the 95\,GHz bandpowers, {\bf black circles} mark the 150\,GHz bandpowers, and the 220\,GHz bandpowers are plotted with {\bf blue diamonds}. 
  Note that the 150\,GHz uncertainties are smaller than the symbols. 
  The \textbf{\textit{top panel}} displays the results of the ``first half - second half" jackknife. 
  The PTE to exceed the measured \chisq\ is 0.90. 
  The  ``left - right" jackknife is shown in the \textbf{\textit{second panel}} and has a PTE of 0.21. 
  The ``azimuth-split" jackknife is plotted in the \textbf{\textit{third panel}} and has a PTE of 0.84. 
  Finally, the auto-spectra of the three frequencies are plotted in the \textbf{\textit{bottom panel}} for comparison.
 }
  \label{fig:jackknife}
\end{figure}

We find the results of each test to be consistent with zero. 
The results are plotted in Figure  \ref{fig:jackknife}. 
Combining all three frequencies, the probabilities to exceed (PTE) for the ``first half - second half", ``left - right" and ``azimuth-split" jackknives are 0.90, 0.21, and 0.84 respectively. 
The PTE of the observed \chisq's is 0.50 for the three tests at 95\,GHz. 
For the four tests at 150\,GHz and 220\,GHz, the PTE  are 0.48 and 0.61 respectively. 
There is no evidence for systematic effects in the SPT data. 

\section{Bandpowers}
\label{sec:bandpowers}

\begin{figure*}[t]\centering
\includegraphics[width=0.95\textwidth]{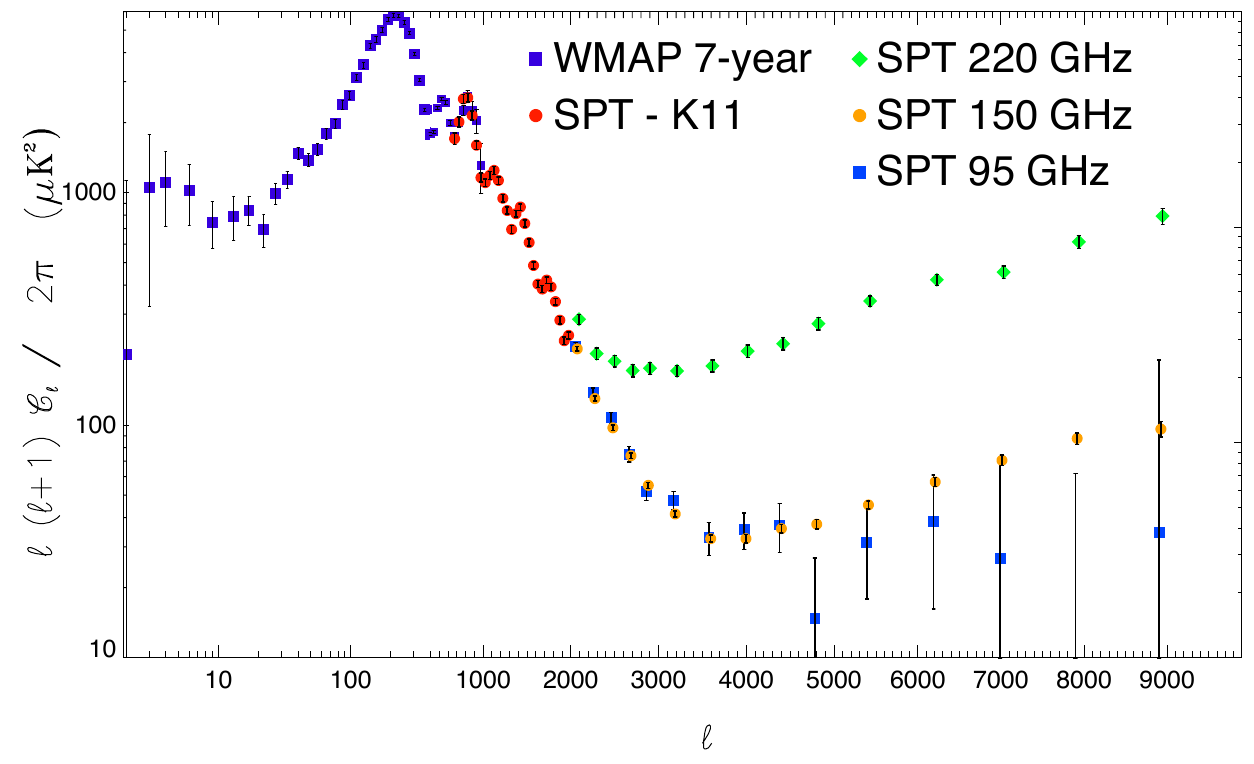}
  \caption[]{The WMAP7 and SPT bandpowers. 
  The SPT bandpowers at $\ell \le 2000$ are taken from K11 and are at 150\,GHz only. 
  At  $\ell \ge 2000$, we show the bandpowers at 95, 150, and 220\,GHz measured with the SPT in this work. 
  Below $\ell = 2000$, the primary CMB anisotropy is dominant at all frequencies. 
  On smaller scales, the CIB, radio sources, and secondary CMB anisotropies contribute to the signal. 
 With the SPT source masking, the CIB is the largest source of power on sub-arcminute scales at 150 and 220\,GHz. 
Due to the relative spectral behavior of the CIB and synchrotron emission, the 95\,GHz bandpowers also have
a significant contribution from radio sources. 
  }
  \label{fig:bandpowersall}
\end{figure*}

Applying the analysis described in \S\ref{sec:bandpowerest} to the 800 \sqdeg\ of sky observed with the SPT in 2008 and 2009 leads to the bandpowers shown in Figure \ref{fig:bandpowersall} and tabulated in Table \ref{tab:bandpowers}. 
The bandpowers, covariance matrix, and window functions are available for download on the SPT 
website.\footnote{http://pole.uchicago.edu/public/data/reichardt11/}

\begin{figure*}[t]\centering
\includegraphics[width=0.9\textwidth]{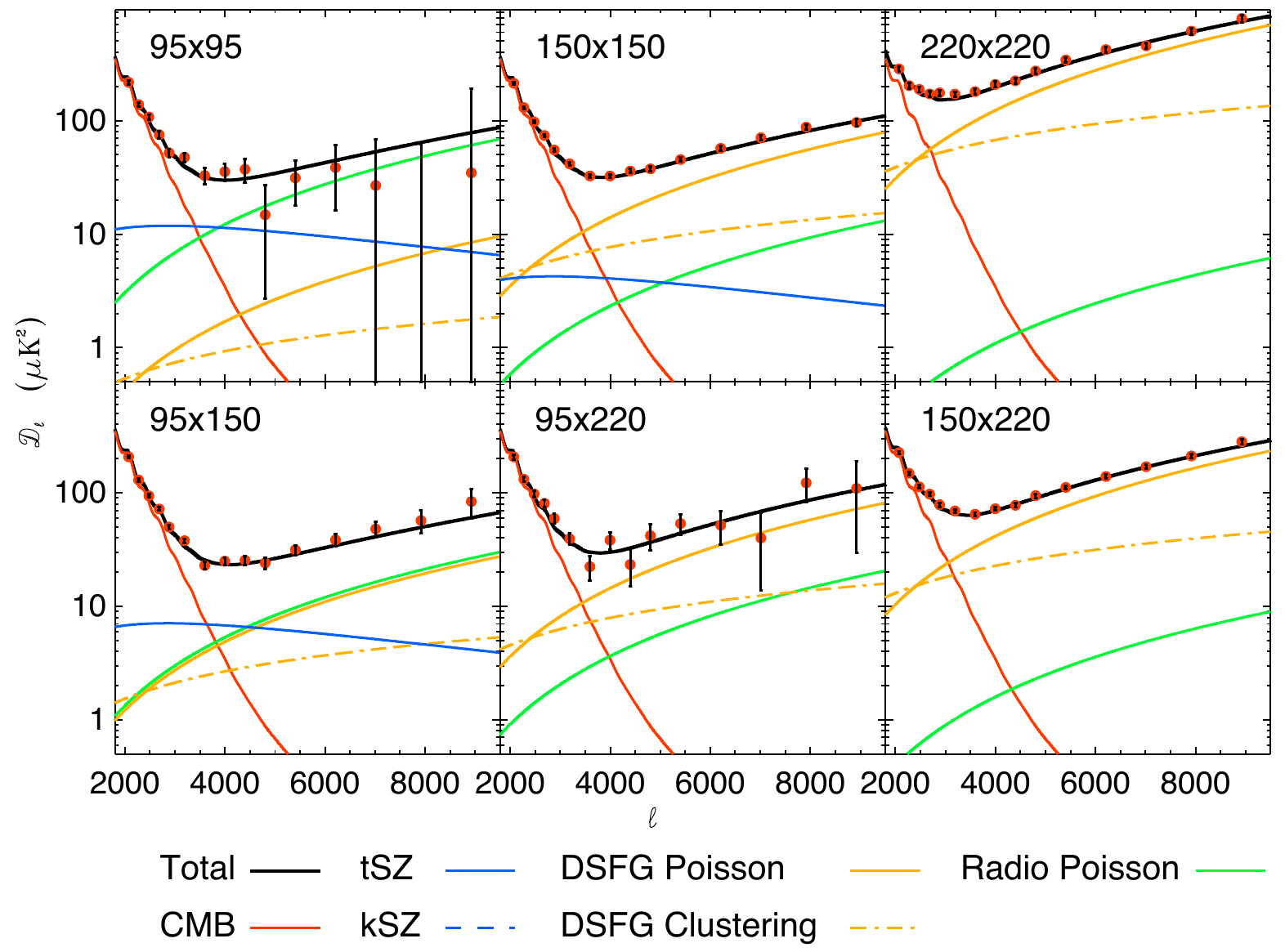}

  \caption[]{ The six auto- and cross-spectra measured with the 3-frequency SPT data. 
  Overplotted on the bandpowers is the best-fit model for the fiducial set of model parameters. 
  The bandpowers have not been corrected by the best-fit calibration or beam uncertainties in the MCMC chains; for reference, the best-fit temperature calibration factors at 95, 150, and 220\,GHz are 0.999, 0.997, and 1.003 respectively. 
  In addition to the complete model ({\bf black lines}), each individual model component is shown.  
  The tSZ effect is marked with the {\bf blue solid  line}. 
  The best-fit kSZ power is near-zero and off-scale. 
  The Poisson power from DSFGs and radio galaxies are shown by {\bf solid orange and green lines} respectively. 
  The clustered  component to the DSFGs is shown with a {\bf orange dot-dash line}. 
}
  \label{fig:bandpowerbestfit}
\end{figure*}

A combination of the primary CMB anisotropy, foregrounds, and secondary SZ anisotropies  is detected at high S/N from $\ell$ = 2000 to 9400 in all three frequency bands (see Figure  \ref{fig:bandpowerbestfit}). 
The largest signal in the maps is the primary CMB anisotropy; at $\ell < 3000$ the bandpowers trace out the Silk damping tail of the CMB power spectrum. 
The next largest signal is the CIB anisotropy which becomes dominant around $\ell = 3000$.  
At high multipoles, the bandpowers follow the expected $\ell^2$ form for Poisson power from point sources. 
The spectral index of the point source contribution is $\alpha \simeq 3.5$ between 150 and 220\,GHz, characteristic of  the Rayleigh-Jeans tail of a dust spectrum. 
Poisson fluctuations of radio sources become important as well at 95\,GHz. 
Over these angular scales, the CMB and Poisson fluctuation power from point sources account for approximately 80\% of the power in all frequency bands. 
We also detect the tSZ effect and clustering component of the CIB  at high significance.

\begin{table*}[ht!]
\begin{center}
\caption{\label{tab:bandpowers} Bandpowers}
\small
\begin{tabular}{cc|cc|cc|cc}
\hline\hline
\rule[-2mm]{0mm}{6mm}
& &\multicolumn{2}{c}{$95\,$GHz} &\multicolumn{2}{c}{$150\,$GHz} & \multicolumn{2}{c}{$220\,$GHz} \\
$\ell$ range&$\ell_{\rm eff}$ &$\hat{D}$ ($\mu{\rm K}^2$)& $\sigma$ ($\mu{\rm K}^2$) &$\hat{D}$ ($\mu{\rm K}^2$)& $\sigma$ ($\mu{\rm K}^2$)&$\hat{D}$ ($\mu{\rm K}^2$)& $\sigma$ ($\mu{\rm K}^2$) \\
\hline

2001 - 2200 & 2073 & 217.4 &   9.1 & 212.9 &   4.8 & 285.2 &  14.3 \\ 
2201 - 2400 & 2274 & 138.4 &   6.4 & 130.1 &   3.1 & 203.0 &  11.3 \\ 
2401 - 2600 & 2479 & 107.5 &   5.8 &  97.4 &   2.5 & 188.3 &  10.8 \\ 
2601 - 2800 & 2685 &  75.1 &   5.7 &  73.9 &   2.0 & 171.7 &  10.5 \\ 
2801 - 3000 & 2883 &  51.9 &   4.5 &  55.1 &   1.7 & 175.6 &  10.5 \\ 
3001 - 3400 & 3190 &  47.5 &   4.3 &  41.5 &   1.2 & 171.2 &   9.1 \\ 
3401 - 3800 & 3594 &  32.8 &   5.3 &  32.5 &   1.1 & 179.8 &  10.3 \\ 
3801 - 4200 & 3996 &  35.5 &   6.3 &  32.5 &   1.3 & 208.2 &  12.4 \\ 
4201 - 4600 & 4400 &  37.2 &   8.9 &  35.9 &   1.5 & 224.0 &  13.9 \\ 
4601 - 5000 & 4803 &  14.8 &  12.1 &  37.5 &   1.8 & 273.3 &  17.0 \\ 
5001 - 5800 & 5406 &  31.3 &  13.4 &  45.4 &   1.8 & 341.8 &  18.3 \\ 
5801 - 6600 & 6209 &  38.6 &  22.4 &  57.0 &   2.5 & 422.0 &  23.3 \\ 
6601 - 7400 & 7012 &  26.8 &  41.7 &  70.6 &   3.6 & 455.1 &  27.9 \\ 
7401 - 8400 & 7917 & -16.3 &  78.3 &  87.7 &   5.0 & 614.0 &  39.8 \\ 
8401 - 9400 & 8924 &  34.7 & 156.0 &  96.2 &   7.3 & 792.5 &  62.1 \\

\hline
&&\multicolumn{6}{c}{}\\
& & \multicolumn{2}{c}{$95\times150\,$GHz} & \multicolumn{2}{c}{$95\times220\,$GHz} & \multicolumn{2}{c}{$150\times220\,$GHz} \\
\hline
2001 - 2200 & 2073 & 207.6 &   5.8 & 207.2 &   9.8 & 226.0 &   7.1 \\ 
2201 - 2400 & 2274 & 129.6 &   3.8 & 131.7 &   7.1 & 148.6 &   4.8 \\ 
2401 - 2600 & 2479 &  94.0 &   3.1 &  97.7 &   6.3 & 112.8 &   3.9 \\ 
2601 - 2800 & 2685 &  72.1 &   2.5 &  80.7 &   6.3 &  97.4 &   3.5 \\ 
2801 - 3000 & 2883 &  49.9 &   2.1 &  59.5 &   5.6 &  78.6 &   3.2 \\ 
3001 - 3400 & 3190 &  37.9 &   1.6 &  39.3 &   4.5 &  69.2 &   2.4 \\ 
3401 - 3800 & 3594 &  22.9 &   1.6 &  22.4 &   5.5 &  64.8 &   2.4 \\ 
3801 - 4200 & 3996 &  25.0 &   1.9 &  38.3 &   6.6 &  72.4 &   2.8 \\ 
4201 - 4600 & 4400 &  25.5 &   2.2 &  23.3 &   8.2 &  77.6 &   3.2 \\ 
4601 - 5000 & 4803 &  24.1 &   2.9 &  42.0 &  10.7 &  94.5 &   4.0 \\ 
5001 - 5800 & 5406 &  31.4 &   3.2 &  53.6 &  10.8 & 111.4 &   4.2 \\ 
5801 - 6600 & 6209 &  38.5 &   4.9 &  51.9 &  17.0 & 139.3 &   5.4 \\ 
6601 - 7400 & 7012 &  48.1 &   7.8 &  40.0 &  26.3 & 169.9 &   7.4 \\ 
7401 - 8400 & 7917 &  57.0 &  13.1 & 122.5 &  39.9 & 211.2 &   9.6 \\ 
8401 - 9400 & 8924 &  83.9 &  24.4 & 109.3 &  79.8 & 282.0 &  15.6 \\ 

\hline
\end{tabular}
\tablecomments{ $\ell$-band multipole range, weighted multipole value $\ell_{\rm eff}$, bandpower $\hat{D}$, 
and bandpower uncertainty $\sigma$ for the six auto and cross-spectra of the $95\,$GHz, $150\,$GHz, and $220\,$GHz maps. 
The quoted uncertainties are based on the diagonal elements of the covariance matrix which includes the instrumental noise, Gaussian sample variance, and beam uncertainties.  
The sample variance of the tSZ effect and calibration uncertainty are not included. 
Calibration uncertainties are quoted in \S\ref{sec:beamcal}.
Point sources with flux $>6.4\,$mJy at $150\,$GHz have been masked out at all frequencies in this analysis. 
This flux cut substantially reduces the contribution of radio sources to the bandpowers, although DSFGs below this threshold contribute significantly to the bandpowers.  }
\normalsize
\end{center}
\end{table*}

\section{Cosmological Modeling}
\label{sec:model}

We fit the SPT bandpowers to a model including lensed primary CMB anisotropy, secondary tSZ and kSZ anisotropies, and foregrounds. 
As stated in the previous section, the CIB is the most important foreground and it is composed primarily of emission from a population of DSFGs. 
The power spectrum of these DSFGs can be divided into ``Poisson" and ``clustered" components. 
Radio sources are also important in the lower frequency bands.  
Finally, we include a template for the galactic cirrus emission which is negligible by virtue of the selected sky coverage.

Parameter constraints are calculated using the publicly available {\textsc CosmoMC}\footnote{http://cosmologist.info/cosmomc} package \citep{lewis02b}.
We have added two modules to handle the high-$\ell$ data: one to model the foregrounds and secondary anisotropies and one to calculate the SPT and \planck\  likelihood functions.
The code is a modified and expanded version of the modules discussed in \citet{millea11} and used in S11.
These modules and instructions for compiling them are available at the SPT website.\footnote{http://pole.uchicago.edu/public/data/reichardt11/}

We include measurements of the primary CMB anisotropy, baryon acoustic oscillations (BAO), and Hubble constant (H$_0$) in all cosmological constraints presented here. 
For the primary CMB, we include both the seven-year WMAP data release (WMAP7, \citealt{larson11}) and the recent large-scale SPT power spectrum (K11). 
We note that the same time-ordered data are used in this work and K11. 
Hence we restrict the K11 data to non-overlapping multipoles ($\ell < 2000$). 
We also subtract a Poisson term from the K11 bandpowers to adjust for the 
different source masking thresholds between the two works (for details see \S\ref{sec:radio}). 
All parameter chains include low-redshift measurements of the Hubble constant $H_0$ using the Hubble Space Telescope \citep{riess11} and the BAO feature using SDSS and 2dFGRS data \citep{percival10}.  
Finally, we include \planck\ CIB measurements at 217\,GHz and in some cases 353\,GHz \citep{planck11-6.6_arxiv}. 
When comparing to the CIB model, we subtract the estimated Poisson radio contribution from the \planck\ CIB bandpowers.

In some MCMC chains, we also compare the measured SZ power with external measurements of low redshift structure. 
We use local galaxy cluster abundances as measured by \citet{vikhlinin09}, which directly and tightly constrain $\sigma_8$ and are consistent with other structure probes \citep{rozo10, mantz10b, vanderlinde10, sehgal11}.  
\citet{vikhlinin09} use the abundance of local ($0.025<z<0.22$) clusters to infer $\sigma_8 (\Omega_m/0.25)^{0.47} = 0.813\pm 0.013\pm 0.024$, where the second set of errors is an estimate of the systematic uncertainty due to the uncertainty in the masses of the clusters. 
This constraint will be referred to as the X-ray $\sigma_8$ prior.

\subsection{Primary Cosmic Microwave Background Anisotropy} 
\label{sec:primarycmb}

We use the standard, six-parameter, spatially flat, lensed 
$\Lambda$CDM cosmological model to predict the primary CMB temperature anisotropy. 
The six parameters are the baryon 
density $\Omega_b h^2$, the density of cold dark matter $\Omega_c h^2$, the optical 
depth to recombination $\tau$, the angular scale of the sound horizon at last scattering $\Theta_s$, the 
amplitude of the primordial density fluctuations $\ln[10^{10} A_s]$, and the 
scalar spectral index $n_s$.

We calculate the primary CMB power spectrum in the MCMC chains using PICO \citep{fendt07a,fendt07b} instead of CAMB \citep{lewis00} for speed. 
 With lensing on, we find chains run with PICO (and the fast WMAP likelihood code \citep{dvorkin10} modified for the latest WMAP7 likelihood) to be roughly 2000 times faster than ones run with CAMB. 
PICO allows us to replace the unlensed spectra plus ``lensing template" approximation used in L10 and S11 with the full lensing treatment. 
We have trained PICO using the January 2011 version of CAMB with lensing including non-linear corrections turned on. 
Note that we observe no change to the foreground or secondary anisotropy constraints from the current data when using the former lensing template approximation. 
However, reducing the lensing power  at $\ell = 3000$ (by turning off lensing or neglecting non-linear effects) would increase the inferred kSZ power by an approximately equal amount. 
For instance, neglecting non-linear effects would increase the kSZ power by $\sim$$\,1\,\mu{\rm K}^2$.

\subsection{Thermal Sunyaev-Zel'dovich Anisotropy}
\label{sec:tsz}

We adopt two different models for the tSZ power spectrum.  S11
considered four tSZ models: a model by \citet{sehgal10} which
predicts high levels of tSZ power and three models which all predicted
 lower tSZ powers \citep{battaglia10, shaw10, trac11}.
Here we have limited the model set to one from each group: the high
power model by \citet{sehgal10} and the model by \citet{shaw10}  from the
lower set.

We use the tSZ power spectrum predicted by \citet{shaw10} as the
baseline model.  \citet{shaw10} investigate the impact of cluster
astrophysics on the tSZ power spectrum using halo model calculations
in combination with an analytic model for the ICM.  These
astrophysical effects -- in particular, the inclusion of non-thermal
pressure support and AGN feedback -- tend to suppress the tSZ power
\citep[see also][]{battaglia11, trac11}.  We use the baseline model
from that work (hereafter the Shaw model), which predicts $D^{\rm
  tSZ}_{3000}=4.3 \,\mu{\rm K}^2$ at $\ell =3000$ and 152.9 GHz.  The
assumed cosmological parameters are ($\Omega_b$, $\Omega_m$,
$\Omega_\Lambda$, $h$, $n_s$, $\sigma_8$) = (0.044, 0.264, 0.736,
0.71, 0.96, 0.80).  We use this model in all MCMC chains where another
model is not explicitly specified.

We also consider the tSZ power spectrum model presented by
\citet[][hereafter the Sehgal model]{sehgal10}.  \citet{sehgal10}
combined the semi-analytic model for the intra-cluster medium (ICM) of
\citet{bode09} with a cosmological N-body simulation to produce
simulated tSZ and kSZ maps from which the template power
spectra were measured.  The assumed cosmological parameters are the
same as with the Shaw model.  At $\ell = 3000$, this model predicts
$D^{\rm tSZ}_{3000}=7.4\, \mu{\rm K}^2$ of tSZ power at 152.9 Ghz. 

As described in S11, the model of \citet{shaw10} is also used to
rescale both tSZ model templates as a function of cosmological
parameters. Around the fiducial cosmological model, the tSZ amplitude roughly scales as
\begin{equation}
D^{\rm tSZ}  \propto \left(\frac{h}{0.71}\right)^{1.7} \left(\frac{\sigma_8}{0.80}\right)^{8.3} \left(\frac{\Omega_b}{0.044}\right)^{2.8}.
\label{eq:tsz_scaling}
\end{equation}

Both tSZ models exhibit a similar angular scale dependence over the
range of multipoles to which SPT is sensitive.  When the normalization
of each model is allowed to vary, we detect similar tSZ power for both
(see Table \ref{tab:szconstraint}).  
However, as we discuss in
\S\ref{sec:sigma8}, the predicted model amplitude is crucial in
interpreting the detected tSZ power as a constraint on $\sigma_8$ and 
also leads to significant changes in the fit quality when the tSZ
power is fixed to cosmologically-scaled model expectations.

\subsection{Kinetic Sunyaev-Zel'dovich Anisotropy}

Thomson scattering between CMB photons and clouds of free electrons that have a
coherent bulk velocity 
produce fluctuations in the observed brightness temperature of the
CMB. This effect, known as the kinetic Sunyaev-Zel'dovich (kSZ) effect,
leads to hot or cold spots, depending on whether the ionized gas is
moving toward or away from the observer. The shift is identical to a change in the CMB blackbody temperature, except for tiny relativistic
corrections.  

KSZ temperature anisotropy requires perturbations in the free electron
density.  This can be due to local perturbations in the baryon
density, $\delta_{\rm b} =\rho_{\rm b}/\bar{\rho}_{\rm b}-1$ (where
$\bar{\rho_b}$ is the mean baryon density in the universe), or  ionization fraction, $\delta_{x} = x_e / \bar{x}_e-1$.
Perturbations that are correlated with the large scale velocity field
will produce an observable change in temperature of the CMB blackbody.
The total contribution to the temperature anisotropy is given by the
integral over conformal time,
\begin{equation}
\frac{\Delta T_{\rm kSZ}}{T_{CMB}} (\n) = \sigma_T\overline{n}_{e,0}
\int d \eta [a^{-2} e^{-\tau(\eta)} \bar x_e(\eta)] \n \cdot \q \;,
\label{eq:ksz}
\end{equation}
where $\sigma_T$ is Thomson scattering cross-section,  $a$ is the scale factor, $\tau(\eta)$ is
the optical depth from the observer to conformal time $\eta$,  
$\bar{n}_{e,0}$ is the mean electron density of the universe at
the present day, and $\n$ is the line of sight unit vector. Finally,  
\begin{equation}
\q=(1+\delta_x) (1+\delta_{\rm b}) \mathbf{v} \, ,
\end{equation}
where ${\bf v}$ represents the peculiar velocity of free electrons at
$\eta$.

The kSZ power spectrum can be broken down loosely into two
components.  The first is the kSZ signal from the post-reionization
epoch, $z < z_{\rm end}$, where we define $z_{\rm end}$ as the
redshift at which hydrogen reionization is complete.  The second
component comes from the epoch of reionization itself.  Models of
inhomogeneous (or patchy) reionization predict bubbles of free
electrons around UV-emitting sources embedded in an otherwise neutral
medium.  
Any large-scale bulk velocity of these bubbles will impart a temperature anisotropy onto the CMB. 
The kSZ signal from
reionization is thus sensitive to fluctuations in both the gas density
and ionization fraction. Note that in the case of instantaneous
reionization, the post-reionization signal is the only source of kSZ
power.

We henceforth refer to the post-reionization component as the
`homogeneous kSZ' signal, and that from the reionization epoch as the
`patchy kSZ'  signal. We now discuss our modeling of each in
more detail.

\subsubsection{Homogeneous Kinetic Sunyaev-Zel'dovich Anisotropy}
\label{sec:hksz}

For the homogeneous (or post-reionization) kSZ signal, we adopt the
{\em non-radiative} (NR) and {\em cooling plus star-formation} (CSF)
models presented by \citet{shaw11}.  These models are constructed by
calibrating an analytic model for the homogeneous kSZ power
spectrum with two hydrodynamical simulations.  The first simulation was
run in the non-radiative regime, while the second included
metallicity-dependent radiative cooling and star-formation.
\citet{shaw11} measured the power spectrum of gas density fluctuations
in both simulations over a range of redshifts, and used this to
calculate the kSZ power spectrum for each case.  They argue that
these two cases are likely to bracket the true homogeneous kSZ power.
For our fiducial cosmology (detailed in \S\ref{sec:tsz}) at $\ell =
3000$, the NR template predicts 2.4 \muksq, whereas the CSF simulation
predicts 1.6 \muksq.  The scaling of these
models with cosmological parameters is given approximately by
\begin{eqnarray}
D^{\rm kSZ}  &\propto& \left(\frac{h}{0.71}\right)^{1.7} \left(\frac{\sigma_8}{0.80}\right)^{4.7} \left(\frac{\Omega_b}{0.044}\right)^{2.1}\times\nonumber\\
&& \left(\frac{\Omega_m}{0.264}\right)^{-0.4} \left(\frac{n_s}{0.96}\right)^{-0.2}.
\label{eq:ksz_scaling}
\end{eqnarray}
Note that, for both models, we assume that reionization occurs
instantaneously at $z_{\rm end} = 8$. 
 Both models also assume that Helium remains neutral; if
Helium is singly ionized at the same time as hydrogen, the
predicted power would increase by 16\%.

\subsubsection{Kinetic Sunyaev-Zel'dovich Anisotropy from Patchy Reionization Scenarios}
\label{sec:pksz}

For the patchy kSZ signal, we adopt the model presented by Z11 and briefly summarized here. 
This model starts from a matter simulation.  
For every position, the
linear matter over-density is calculated within some radius. 
The halo collapse fraction is estimated from this over-density, and
translated into an expected number of ionizing photons based on the
number of collapsed halos above the atomic cooling threshold mass.  
The number of ionizing photons is then compared to the number of hydrogen atoms
within this sphere.  
If there are sufficient ionizing photons, the
sphere is labeled ionized.  If not, a smaller  radius is set
and the algorithm repeated until the resolution of the simulation box is reached. 
The resulting ionization field is used in conjunction with the underlying density and velocity fields to
compute the patchy kSZ power spectrum.

As discussed in Z11, the detailed reionization history used for the kSZ template is unimportant, since the template {\em shape} is robust to the duration and the mean redshift of reionization. 
The amplitude of the patchy kSZ signal is nearly proportional to the duration of reionization (with a mild redshift dependence), making it an excellent probe of reionization.
The kSZ effect is complementary
 to the large scale CMB polarization anisotropy (the reionization bump) that provides a constraint on the global timing of reionization, and to QSO and Ly$\alpha$ emitter observations that study the very end of reionization. 
For the fiducial model template provided by Z11 and used in this work, reionization begins at $z=11$ and concludes at $z=8$.

\subsection{Cosmic Infrared Background}
\label{sec:cib}

The CIB is produced by thermal emission
from DSFGs over a very broad range in
redshift \citep{lagache05,marsden09}.  The dust grains, ranging in
size from a few molecules to 0.1 mm, absorb light at wavelengths
smaller than their size, and re-radiate it at longer wavelengths. 
The result of this absorption is roughly equal amounts of energy in the CIB as in
 the unprocessed starlight that makes up the
optical/UV background \citep{dwek98,fixsen98}.

The frequency dependence of the CIB (and radio galaxies in \S\ref{sec:radio}) is
naturally discussed and modeled in units of flux density (Jy) rather
than CMB temperature units.  However, as the power spectrum is
calibrated in CMB temperature units, we need to determine the ratio of
powers between different frequency bands in CMB temperature units.  
To calculate the ratio of power in
the $\nu_i \times \nu_j$ cross-spectrum to the power at (the
arbitrary frequency) $\nu_0$ in units of CMB temperature squared, we
multiply the ratio of flux densities by
\begin{equation}
\epsilon_{\nu_1,\nu_2} \equiv \frac{\frac{dB}{dT}|_{\nu_0} \frac{dB}{dT}|_{\nu_0}}{\frac{dB}{dT}|_{\nu_i} \frac{dB}{dT}|_{\nu_j}},
\end{equation}
where $B$ is the CMB blackbody specific intensity evaluated at $T_{\textrm{CMB}}$, and $\nu_i$ and $\nu_j$ are the effective frequencies of the SPT bands.  Note that $\nu_i$ may equal $\nu_j$.
The effective frequencies of the SPT bandpasses for each foreground are presented in \S\ref{subsec:efffreq}.

\subsubsection{Poisson term}
\label{sec:cibpoisson}

Statistical fluctuations in the number of DSFGs lead to a Poisson distribution on the sky. 
DSFGs are effectively point sources at these angular scales, so they contribute a constant ${\it C}_\ell$ (thus ${\it D}_\ell \propto \ell^2$) to the SPT bandpowers. 
The bulk of the millimeter-wavelength CIB anisotropy power is expected to come from faint ($\lesssim 1\,$mJy) DSFGs - so the residual CIB power is largely independent of the source mask flux threshold. 
This is a key assumption in combining the \planck\ and SPT data as the flux cuts differ by a factor of eight at 220\,GHz. 
The 217\,GHz \planck\ cut is at 160\,mJy while the 150\,GHz SPT flux cut of 6.4\,mJy scales to approximately 22\,mJy at 220\,GHz. 
We test the impact of the flux cut with the modeled source counts of \citet{bethermin11} and \citet{negrello07}, and find the two models predict a difference in the Poisson power of 4\% or 16\% respectively. 
Hence in the model fitting, we scale the modeled DSFG Poisson power up by 10\% when comparing the model to the  \planck\ CIB bandpowers.
We have tested changing this assumption by tens of percent and find the cosmological constraints are insensitive to this -- the Poisson term is subdominant at all multipoles ($\ell \in [80,2240]$) in the \planck\ bandpowers. 

The Poisson power 
spectrum of the DSFGs can be written as
\begin{equation}
D^{p}_{\ell,\nu_1,\nu_2} = \amplitudeletter{p} \epsilon_{\nu_1,\nu_2} \frac{\eta_{\nu_1} \eta_{\nu_2} } {\eta_{\nu_0} \eta_{\nu_0}}  \left( \frac{\ell}{3000}\right)^2,
\end{equation}
where $\eta_{\nu} $ encodes how the CIB brightness scales with observing frequency (see  \S\ref{subsec:freqdep}). 
$\amplitudeletter{p}$ is the amplitude of the Poisson DSFG power spectrum at $\ell=3000$ and frequency $\nu_0$. 
Note that we neglect the decohering effect of frequency dependence varying from source to source.  
\citet{hall10} argued that the  spectral index variation in the DSFG population was less than $\sigma_\alpha=0.7$. 
This level of variation has a negligible effect on the inferred Poisson powers across the SPT frequency bands.

\subsubsection{Clustered term}
\label{sec:cibcluster}

DSFGs (and galaxies in general) trace the mass distribution, so they are spatially correlated. 
This leads to a ``clustering'' term in the DSFG power spectrum in addition to the Poisson term. 
On the physical scales of relevance, galaxy correlation functions are surprisingly well-approximated by a power law.  
The clustering shape we consider is thus a phenomenologically-motivated power law with index chosen to match the correlation properties observed for Lyman-break galaxies at $z \sim 3$ \citep{giavalisco98,scott99}.  
This form has been used previously by a number of millimeter- and submillimeter-wavelength experiments (\citet{viero09, dunkley11, planck11-6.6_arxiv, addison11}, S11). 
We adopt this ``power law'' model to represent the non-linear regime, and use a clustering template of the form $\shapeletter{c} \propto \ell^{0.8}$ in all cases unless specified.

In \S\ref{subsec:powerlaw}, we consider two alternate forms for the clustering term. 
First, we free the power-law exponent, i.e., $\shapeletter{c} \propto \ell^{\gamma}$ with $\gamma$ free. 
Second, we add the linear-theory model by \citet{hall10} with a free amplitude to the power-law clustering term. 
The fiducial model in that work was calculated assuming that light is a biased tracer of mass fluctuations calculated in linear perturbation theory, with redshift and scale--independent bias.  
Further, the model assumed all galaxies have the same greybody SED, and the redshift distribution of the luminosity density was given by a parametrized form, with parameters adjusted to fit SPT, BLAST, and 
{\it Spitzer} power spectrum measurements.  
We find that the angular template shape depends only slightly on the observing frequency and choose to use the 150\,GHz model spectrum. 
Here we refer to this power spectrum shape as the ``linear theory'' template.
As will be discussed in \S\ref{sec:cibconstraint}, the data strongly prefer the power-law form over the linear theory clustering template. 
The data are also consistent with the assumed power law exponent of $\gamma=0.8$.

\subsubsection{Frequency dependence}
\label{subsec:freqdep}

Unless otherwise noted, all results presented in this work assume the modified black-body (modified BB) model  for the frequency dependence of the CIB: 
\begin{equation}
\eta_{\nu}  = \nu^\beta B_\nu(T),
\end{equation}
where $B_\nu(T)$ is the black-body spectrum for temperature T, and $\beta$ is an effective dust emissivity index. 
T and $\beta$ are free parameters with uniform priors $T \in [ 5,35 \,K]$ and $\beta \in [0,2]$. 
The upper limit on the temperature prior is unimportant because the SPT frequencies are in the Rayleigh-Jeans tail of $>$15 K objects. 
Therefore, there would be a perfect degeneracy between 30~K and 100~K dust. 
We have tested instead fixing the temperature to 9.7\,K with a wider $\beta$ range as was done in \citet{addison11} and found little effect on the results reported in this work.

For some results, we also examine the single-SED model presented by \citet{hall10} (hereafter the Single SED model); however, we do not find substantive differences in the results from the modified BB model.
This model assumes a single SED for all DSFGs. 
This SED is a modified black-body with $T = 35 $~K and $\beta = 2$. 
The DSFGs have a density as a function of redshift parametrized by a Gaussian
 with mean \zcenter \ and variance  $\sigma_z^2$. 
The number counts are set to zero at $z < 0$. 
We fix all but  one parameter in the  model; we allow \zcenter\ to vary freely.
As in \citet{hall10}, we set $\sigma_z = 2$. 
The frequency dependence predicted by the Single SED model has a slight $\ell$-dependence; we ignore this and use the frequency scaling at $\ell=3000$. 
We use the model at $\ell = 3000$; it has a slight $\ell$-dependence which we ignore. 
The Single SED model has one fewer parameter and is more easily interpretable physically than the modified BB model. 
However, the data are in mild tension with the model - the preferred \zcenter\ is near the prior cutoff at 0. 
This does not degrade the quality of the fit to the data, but it leads to a one-sided distribution of CIB spectral indices which results in a slight artificial reduction in some parameter uncertainties.

A simplifying assumption we make for the analyses presented here is that the various components of the CIB (Poisson and clustering terms) have the same frequency dependence.  Even in the context of the Single SED model, we expect their spectral indices to differ to some degree, because the weighting of sources as a function of redshift will be different for the two terms.  Our assumption here is, effectively, 
that the redshift distributions of the source populations giving rise to the
  Poisson and clustered components are not sufficiently different as to cause a
  significant difference in frequency dependence.
We have tested allowing separate frequency dependencies for the Poisson and clustered CIB terms and find minimal impact on the SZ constraints.

We have also checked whether the higher frequency, 217 and 353\,GHz \planck\ CIB bandpowers  \citep{planck11-6.6_arxiv} are consistent with the assumption of a single frequency dependence for the
Poisson and clustered terms. 
If the frequency dependence of each term was different, then we would expect to see variations in the CIB power ratio between 217 GHz and 353 GHz as a function of angular scale since the relative power in each term is $\ell$-dependent. 
We find that the CIB power ratio is consistent with a single spectral index with $\alpha_{217/353} = 3.22\pm 0.08$. 
With (9-1) dof, the  null hypothesis of a single spectral index has $\chi^2 = 8.0$ -- the probability to exceed this \chisq\ is 0.44. 
This spectral index is flatter than preferred by  the lower frequency SPT data; however, we expect the CIB frequency dependence to flatten towards higher frequencies as the rest frame emission frequency approaches the peak of the grey body dust emission.

\subsubsection{tSZ-CIB correlations}
\label{subsec:tszcib}

As tracers of the same dark matter distribution, we expect some degree
of correlation between the tSZ and CIB power spectra.  In previous papers,
we have argued that the correlation is likely to be small since most
of the DSFGs contributing to the CIB are in the
field rather than in the massive galaxy clusters that contribute most
of the tSZ power.  Unfortunately, observational constraints on
the IR flux (and therefore tSZ-CIB correlation) are largely limited to 
low-redshift, high-mass clusters.  The observations are also at
frequencies close to the peak of the CIB emission, introducing a
significant scaling uncertainty in translating the results to
the longer wavelengths important for tSZ surveys. The simulations of
\citet{sehgal10} show that a model with significant correlations at
148 GHz 
can be constructed without violating current observational constraints by concentrating the CIB galaxies in the lower-mass, higher-redshift galaxy clusters.

We adopt a simplified treatment of tSZ-CIB correlations in this work.  We
assume the correlation, $\xi$, is independent of angular multipole, and that both the Poisson and clustering CIB components
 correlate with tSZ fluctuations. 
 Note that
the choice of which CIB terms correlate does not significantly change
the resulting tSZ and kSZ constraints, although it does rescale the
amplitude of the inferred correlation.

The tSZ-CIB correlation introduces a term to the model of the form:
\begin{equation}
\label{eqn:tszcib}
D^{tSZ-CIB}_{\ell,\nu_1,\nu_2} = \xi \left( \sqrt(D^{\rm tSZ}_{\ell,\nu_1,\nu_1}  D^{\rm CIB}_{\ell,\nu_2,\nu_2} )  +   \sqrt(D^{\rm tSZ}_{\ell,\nu_2,\nu_2}  D^{\rm CIB}_{\ell,\nu_1,\nu_1})   \right).
\end{equation}
Here $D^{tSZ-CIB}_{\ell,\nu_1,\nu_2}$ is the power due to correlations (which should be negative 
below the tSZ null of 217\,GHz, i.e. $\xi < 0$), $D^{\rm tSZ}_{\ell,\nu_1,\nu_1}$ is the tSZ power spectrum, and $D^{\rm CIB}_{\ell,\nu_1,\nu_1}$ is the sum of the Poisson and clustered CIB components. 

\subsection{Radio galaxies}
\label{sec:radio}

High-flux point sources in the SPT maps are coincident with known radio sources from long-wavelength catalogs (e.g. SUMSS, \citealt{mauch03}), and have spectral indices consistent with synchrotron emission  \citep{vieira10}. 
The Poisson radio source term  can be expressed as:
\begin{equation}
D^{r}_{\ell,\nu_1,\nu_2} = \amplitudeletter{r} \epsilon_{\nu_1,\nu_2} \eta_{\nu_1,\nu_2} ^{\alpha_{r} + 0.5 {\rm ln}(\eta_{\nu_1,\nu_2}) \sigma_{r}^2}
 \left( \frac{\ell}{3000}\right)^2,
\end{equation}
where $\eta_{\nu_1,\nu_2}  = (\nu_1 \nu_2 / \nu_0^2)$ is the ratio of the frequencies of the spectrum to the base frequency. 
 $\amplitudeletter{r}$ is the amplitude of the Poisson radio source power spectrum at $\ell=3000$ and frequency $\nu_0$. 
The mean spectral index of the radio source population is $\alpha_r$ with scatter $\sigma_r$.

The radio source power ($\amplitudeletter{r}$) in the map depends approximately linearly on the flux above which discrete sources are masked. 
This reflects the fact that $S^2 dN/dS$ for synchrotron sources is nearly independent of $S$ (e.g. \citet{dezotti05}).  
As a result, the Poisson radio source power in the K11 bandpowers (which had a higher flux cutoff) is expected to be $9.2\, \pm 2.4\, \mu{\rm K}^2$ higher than in the 150\,GHz bandpowers presented here. 
We correct for this by subtracting a Poisson term with amplitude $\amplitudeletter{}= 9.2\,  \mu{\rm K}^2$ from the K11 bandpowers in the parameter fits. 
Note that the restriction to $\ell < 2000$ we impose on the K11 bandpowers renders the uncertainty on the Poisson power insignificant. 
We have tested varying the subtracted Poisson power and found no impact on the results. 

Following the treatment in S11, we apply a strong prior to the Poisson radio source term due to sources below the SPT detection threshold. 
First, we fix the mean spectral index ($\alpha_r = -0.53$) and scatter ($\sigma_r = 0$). 
Second, based on the \citet{dezotti05} source count model, we apply a Gaussian prior to the 150\,GHz amplitude of $1.28 \pm 0.19 \,\mu{\rm K}^2$. 
We have tested the impact of removing the radio priors.  
The data prefer parameter values consistent with the priors, and there are no significant changes to the CIB and SZ constraints. 
The strong radio priors are included in all cases presented here.

As discussed in \citet{hall10}, the clustering of radio galaxies at SPT frequencies is likely to be negligible.  
Clustering on scales of a few arcminutes is a $\sim5$\% modulation of the mean,
and the intensity of the radio source mean is quite low.  
Furthermore in the unified model of active galactic nuclei, the dominant sources at $>$\,90\,GHz (blazars and flat spectrum radio quasars) are active galactic nuclei with their jets aligned towards the observer. 
The effect of random alignment would serve to further dilute the clustering signal. 
Supporting this argument are conservative extrapolations of upper limits on arcminute scale power at 30\,GHz \citep{sharp10}.
If entirely due to radio point source power, the 30\,GHz limit translates to an upper limit of 0.1 $\mu$K$^2$ at 150\,GHz.  
We do not include a clustered radio component in any results in this work. 

\subsection{Galactic Cirrus}
\label{subsec:cirrus}

Galactic cirrus is the final (and smallest) foreground we include in our modeling. 
Following the cirrus treatment in \citet{hall10} and K11, we cross-correlate the SPT maps with the model 8 galactic dust predictions of \citet{finkbeiner99}. 
Galactic cirrus is detected in all frequency bands at a significance of $\sim 3\,\sigma$ with an angular dependence of:
\begin{equation}
D^{\rm cir}_{\ell,\nu_1,\nu_2} = \amplitudeletter{{\rm cir}, \nu_1,\nu_2} \left( \frac{\ell}{3000}\right)^{-1.2}.
\end{equation}
The measured powers are 0.16, 0.21, and $2.19\,\mu{\rm K}^2$ at 95, 150, and 220\,GHz. 
Note that the 150\,GHz amplitude is different from the value quoted in K11 because of the different field weights in the two analyses. 
We fix the amplitude of the galactic cirrus in each frequency band; allowing it to vary with a prior based on the measurement uncertainty  has no impact on any parameter constraints.

\subsection{Effective frequencies of the SPT bands}
\label{subsec:efffreq}

Throughout this work, we refer to the SPT frequencies as 95, 150, and 220\,GHz.
The data are calibrated to CMB temperature units, hence the effective frequency is irrelevant for a CMB-like source.
However, the effective band center will depend (weakly) on the source spectrum for other sources.
The actual SPT spectral bandpasses are measured by Fourier transform spectroscopy, and the effective band center frequency is calculated for each source, assuming a beam-filling source with a nominal frequency dependence. 
The bandpasses differ slightly from 2008 to 2009 due to the focal plane refurbishment. 
We calculate the effective frequency for each year's spectral bandpass and average them according to the yearly weights presented in \S\ref{subsec:data}. 
For an $\alpha = -0.5$ (radio-like) source spectrum, we find band centers of 95.3, 150.2 and $214.1\,$GHz.
For an $\alpha = 3.5$ (dust-like) source spectrum, we find band centers of 97.9, 153.8 and $219.6\,$GHz.
The band centers in both cases drop by $\sim$$\,0.2\,$GHz if we reduce $\alpha$ by 0.5.
We use these radio-like and dust-like band centers to calculate the frequency scaling between the SPT bands for the radio and DSFG terms.
For a tSZ spectrum, we find band centers of 97.6, 152.9 and $218.1\,$GHz.
The ratio of tSZ power in the 95\,GHz band to that in the $150\,$GHz band is 2.78. 
The 220\,GHz band center is effectively at the tSZ null; the ratio of tSZ power in the  220\,GHz band to that in the $150\,$GHz band is 0.00004 for a non-relativistic tSZ spectrum. 

\section{Results}
\label{sec:results}

\subsection{Baseline Model Results}
\label{sec:baseresults}
We adopt a baseline model with the six $\Lambda$CDM parameters and six additional parameters for the secondary CMB anisotropies and foregrounds described in \S\ref{sec:model}. 
Four of these extra parameters represent the power in the tSZ effect, kSZ effect, DSFG Poisson term, and DSFG clustering term.
The final two parameters describe the frequency dependence of the DSFG anisotropy. 
All DSFG terms are assumed to have the same frequency scaling, based on the modified BB model. 
As templates for the angular shape of tSZ and kSZ power spectra, we use the \citet{shaw10} tSZ model and \citet{shaw11} CSF kSZ model.   
The DSFG Poisson power (like all unresolved Poisson power) has the shape $D_\ell \propto \ell^2$ while the 
DSFG clustering term is described by the power-law $D_\ell \propto \ell^{0.8}$ template presented in \S\ref{sec:cibcluster}. 
Figure  \ref{fig:bandpowerbestfit} plots the best-fit values for each component on top of the SPT bandpowers.

Table~\ref{tab:deltachisq} shows the improvement in the quality of the fits with the sequential
introduction of free parameters to the original $\Lambda$CDM primary CMB model.
 Adding additional free parameters beyond the tSZ amplitude does not significantly improve the quality of the fits. 
 However, small improvements are seen with the introduction of tSZ-CIB correlations or treating the clustered and Poisson frequency scalings as independent.

\begin{table}[ht!]
\begin{center}
\caption{\label{tab:deltachisq} Delta $\chi^2$ for model components}
\small
\begin{tabular}{cc|cc}
\hline\hline
\rule[-2mm]{0mm}{6mm}
term & dof & \delchisq\  \\
\hline
CMB&6 & (reference)\\
DSFG Poisson&3 &-1645 \\
DSFGs NL Clustering&1&-218\\ 
Radio Poisson&0 & -164\\
tSZ & 1 & -55 \\
 kSZ & 1 & 0 \\
\hline
\hline
Single SED & -1 & 0 \\
  $\alpha_p \ne \alpha_c$ & 2 & -4\\
    tSZ-CIB correlation & 1 & -2 \\
 $\ell ^\gamma$ clustered DSFGs & 1 & 0 \\
 LT clustered DSFGs & 1 & 0 \\
 
\hline
\end{tabular}
\tablecomments{ Improvement to the best-fit $\chi^2$ as additional terms are added to the model. 
Terms above the double line are included in the baseline model, with each row showing the improvement in likelihood relative to the row above it. 
For rows below the double line, the $\Delta\chi^2$ is shown relative to the baseline model rather than the row above it. 
Here we have assumed the modified BB frequency scaling of the CIB except for the row labeled ``Single SED" which assumes the alternate Single SED frequency scaling.  
The row labeled ``$\alpha_p \ne \alpha_c$" allows the Poisson and clustered power from DSFGs to have independent frequency scalings. 
The  row labeled ``tSZ-CIB correlation" introduces a correlation coefficient between the tSZ and CIB power spectra. 
The last two rows add freedom to the clustered DSFG model shape, either by varying the power-law exponent (``$\ell ^\gamma$ clustered DSFGs") or adding a free amplitude linear theory template (``LT clustered DSFGs").
} \normalsize
\end{center}
\end{table}

The additional parameters discussed in \S\ref{sec:model} are either fixed or constrained by strong priors. 
A Poisson population of radio galaxies is included with a fixed spectral index $\alpha_{r}=-0.53$, zero 
intrinsic scatter, and an amplitude of $D^r_{3000}= 1.28 \pm 0.19 \,\mu{\rm K}^2$. 
Weakening these radio priors -- removing the amplitude prior and setting a uniform prior on the radio spectral index of $ \alpha_{r} \in [-1.5, 0]$ -- does not change the inferred SZ levels. 
We always include the constant Galactic cirrus power described in \S\ref{subsec:cirrus}. 
We assume no correlation between the tSZ effect and CIB in the baseline model. 
This assumption is relaxed in \S\ref{sec:tszcibbaseresults}. 
The \chisq\ for the 90 SPT bandpowers (84 dof since the $\Lambda$CDM parameters are essentially fixed by external data and there are six model parameters beyond $\Lambda$CDM) is 73.3; the PTE for this $\chi^2$ is 79\%. 
We have tested fixing the $\Lambda$CDM parameters parameters to the best-fit values, and find the derived constraints on the six extended model parameters are  unchanged. 
This baseline model fits the SPT data well and provides the most concise interpretation of the data.

\begin{table*}[ht!]
\begin{center}
\caption{\label{tab:szconstraint} SZ constraints}
\small
\begin{tabular}{ccc|c| c}
\hline\hline
\rule[-2mm]{0mm}{6mm}
tSZ Model & kSZ Model & CIB model & $D^{\rm tSZ}_{3000} ~(\mu {\rm K}^2)$ & $D^{\rm kSZ}_{3000} ~(\mu {\rm K}^2)$  \\
\hline

Shaw &  CSF & modified BB & $3.65 \pm 0.69$ & $<2.8$\\

Shaw &  CSF & modified BB w. correlations & $3.26 \pm 1.06$ & $<6.7$\\

Shaw &  patchy & modified BB w. correlations & $3.66 \pm 0.86$ & $<5.7$\\

Shaw &  {\it fixed} CSF & modified BB & $3.34 \pm 0.58$ & (1.57) \\
\hline
\end{tabular}
\tablecomments{ Measured tSZ power and 95\% confidence upper limits on the kSZ power at $\ell=3000$ in   the SPT 150\,GHz band. 
  Two tSZ model templates (the Shaw and Sehgal
  models) and two kSZ models (the CSF homogeneous kSZ model and the patchy reionization kSZ model) 
    have been considered. 
    In most cases, we show results only for the Shaw tSZ template and CSF kSZ template as the other cases are essentially identical -- 
  indicating the data are insensitive to the differences in the modeled angular dependencies.  
  In all cases, the modified BB model is assumed for the CIB frequency scaling. 
  As an extension to the CIB model, we present constraints when correlations are allowed between the tSZ and CIB. 
  This correlation allows twice the kSZ power and weakens the tSZ power constraint by 50\%. 
    The results depend somewhat on the kSZ template when tSZ-CIB correlations are introduced. 
      The real kSZ spectrum should be an mixture of the two kSZ templates. 
  Finally, we show
  the tSZ constraint derived when fixing the kSZ power to that
  predicted by the CSF homogeneous kSZ model.  The CSF kSZ
  model predicts $1.57\,\mu{\rm K}^2$ for the WMAP7 cosmology detailed
  in \S\ref{sec:hksz} and is scaled with cosmology according to the
  prescription in that section.  } \normalsize
\end{center}
\end{table*}

\begin{figure*}[t]\centering
\includegraphics[width=0.9\textwidth]{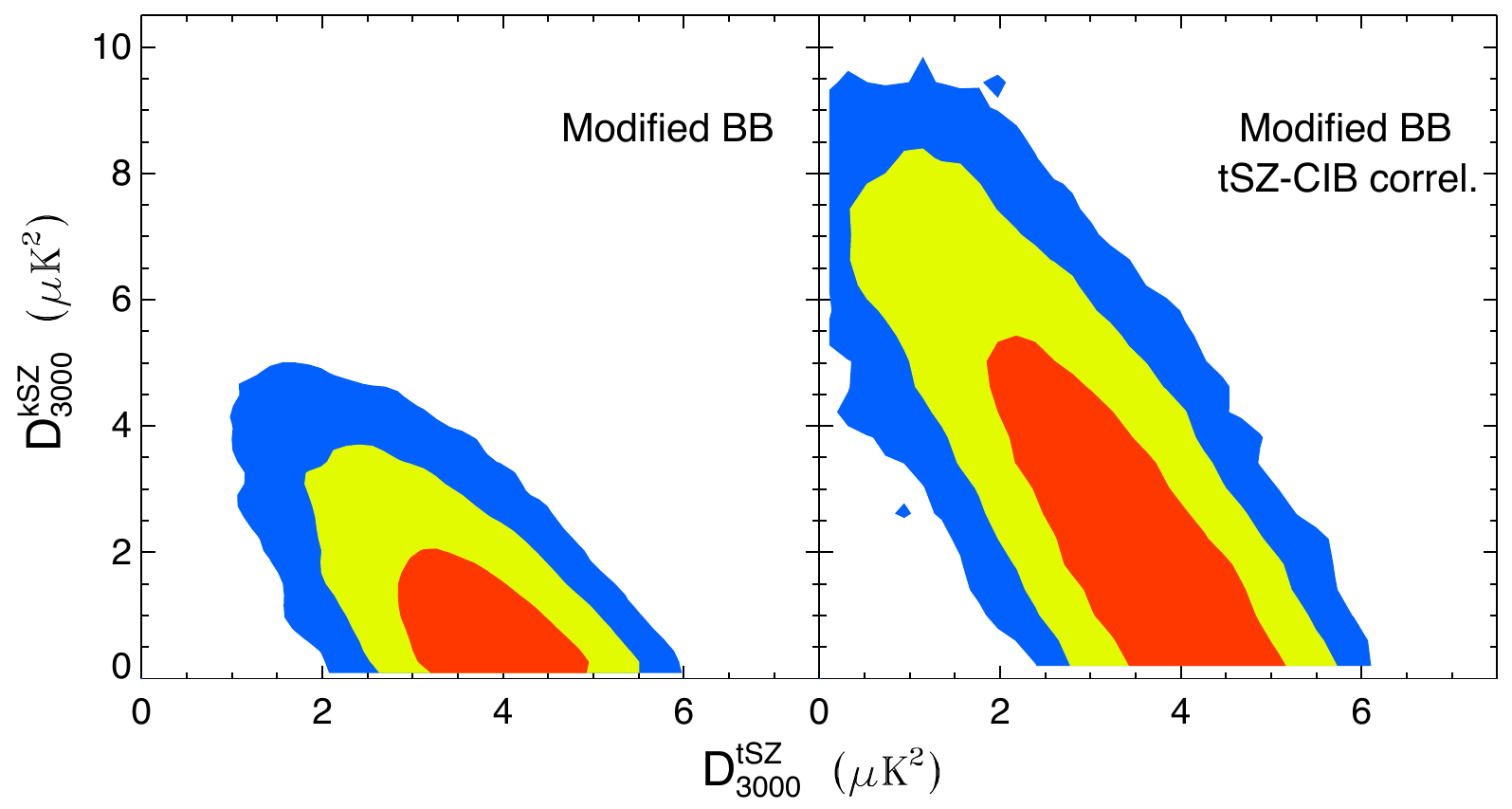}

  \caption[]{ 2D likelihood surface for the tSZ and kSZ power in the SPT 150\,GHz band at $\ell=3000$. 
  1, 2, and 3-$\sigma$ constraints are shown in red, yellow, and blue respectively. 
  The \textbf{\textit{left panel}} shows the constraints for the modified BB CIB frequency scaling. 
    The \textbf{\textit{right panel}} displays the likelihood surfaces for the modified BB CIB frequency scaling with the introduction of free parameter, $\xi$, describing the correlations between the tSZ and CIB power spectra. The tSZ-CIB correlation is degenerate with the kSZ effect, thereby weakening the kSZ limits. 
}
  \label{fig:2dtszksz}
\end{figure*}

SZ parameter constraints for the baseline model and extensions are
listed in Table \ref{tab:szconstraint} and are presented in Figure
\ref{fig:2dtszksz}.  For the tSZ power, we find $D_{3000}^{\rm tSZ}=3.65
\pm 0.69 \,\mu{\rm K}^2$, which is consistent with the predictions of
the Shaw model.  We find a strong 95\% CL upper-limit on kSZ power at
$D_{3000}^{\rm kSZ} < 2.8 \,\mu{\rm K}^2$, although we stress that this
limit is dependent on our DSFG modeling assumptions.  We have applied
a positivity prior to the kSZ power; without this prior, the median
kSZ power is negative by 1.2\,$\sigma$.  
The SZ constraints are
independent of the assumed template shape;  we find similar power
constraints for all tSZ and kSZ model templates considered.

The current tSZ and kSZ constraints are consistent with earlier measurements of the SZ power. 
S11 measured $D_{3000}^{\rm tSZ} + 0.5 D_{3000}^{\rm kSZ} = 4.5 \pm 1.0 \,\mu{\rm K}^2$.  
The equivalent constraint from these data is $(1.034) D_{3000}^{\rm tSZ} + 0.5 D_{3000}^{\rm kSZ} = 4.27 \pm 0.58 \,\mu{\rm K}^2$. 
The multiplicative factor in parentheses before the tSZ term corrects for the different effective bandcenters in the two datasets. 
\citet{dunkley11} measured $D_{3000}^{\rm tSZ} +  D_{3000}^{\rm kSZ} = 6.8 \pm 2.9 \,\mu{\rm K}^2$, while these data prefer $(1.17) D_{3000}^{\rm tSZ} +  D_{3000}^{\rm kSZ} = 5.27 \pm 0.76 \,\mu{\rm K}^2$.
We note that the two analyses differ in that \citet{dunkley11} fixed the ratio of thermal and kinetic SZ power, while this ratio is free in this work. 
The three datasets show excellent consistency. 

We detect both the DSFG Poisson and clustering power with high significance, with powers of $D^{p}_{3000}=7.54 \pm 0.38\, \mu {\rm K}^2$ and $D^{c}_{3000}=6.25 \pm 0.52 \,\mu{\rm K}^2$ respectively at 150\,GHz. 
The  effective spectral index from 150 to 220\,GHz of the DSFG power is found to be $\alpha=3.56 \pm 0.07$. 
The DSFG constraints are discussed in more detail in \S\ref{sec:cibconstraint}, and are in line with both theoretical expectations and previous work (H10; S11; \citealt{dunkley11}).

\subsection{Results for Alternative Models}

Here we explore modifications to the baseline model and the impact on SZ and CIB constraints. 
First, we consider allowing a correlation between the tSZ effect and CIB, and show the impact on the derived SZ power. 
The SZ interpretation is robust to the other CIB model variants we examined. 
Second, we examine model extensions to the $\Lambda$CDM cosmology, and show they do not affect the SZ and CIB parameters.

\subsubsection{ tSZ-CIB Correlations}
\label{sec:tszcibbaseresults}

\begin{figure*}[thb]\centering
\includegraphics[width=0.9\textwidth]{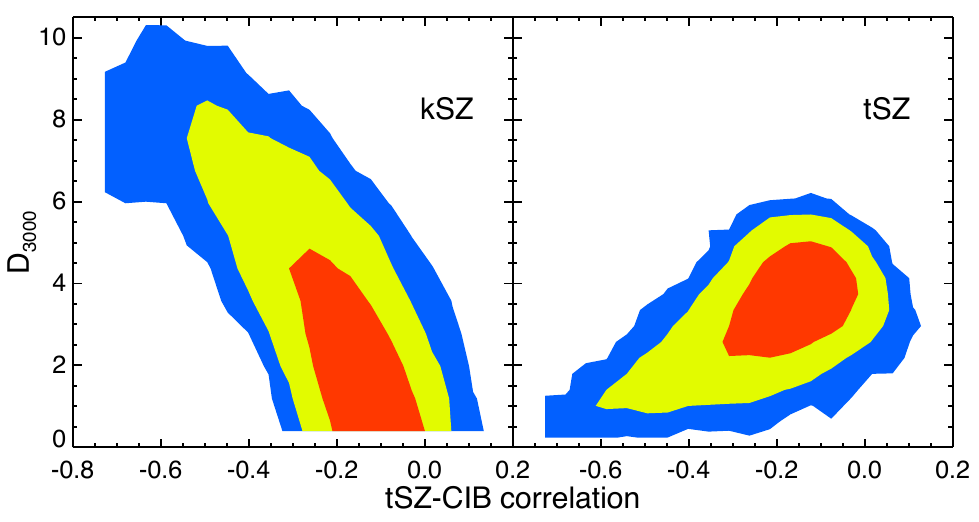}
  \caption[]{ 2D Likelihood curves for the correlation between the tSZ and CIB spectra versus the kSZ power (\textbf{\textit{Left panel}}) or tSZ power (\textbf{\textit{Right panel}}) in $\mu {\rm K}^2$. 
  The filled contours show the 1, 2, and $3\,\sigma$ constraints. 
  The kSZ template consists of the sum of CSF homogeneous kSZ model and patchy kSZ contribution for a model where reionization begins at $z = 11$ and ends at $z=8$. 
  Increasing anti-correlation increases the allowed kSZ power and reduces the tSZ power. 
  The data prefer anti-correlation suggesting that DSFGs are over-dense in galaxy clusters. 
  }
  \label{fig:2dkszcorrel}
\end{figure*}

We first consider the effect of a correlation between the tSZ and CIB
as prescribed by equation \ref{eqn:tszcib}.  Note that DSFG
over-densities in galaxy clusters would lead to an anti-correlation since
the tSZ effect causes a flux decrement at the SPT frequencies.  The
correlated power is highly degenerate with the kSZ and tSZ power as shown in
Figure \ref{fig:2dkszcorrel}.  
We see some dependence on the kSZ template
shape (not shown); the patchy template, which is falling instead of
rising slightly across the angular multipoles of interest, is less
degenerate with the correlation coefficient than the CSF homogeneous
kSZ model.
Increasing anti-correlation increases 
the allowed kSZ power, while reducing the tSZ power, until $\xi \sim -0.4$.  
The slope of the trade-off
between $\xi$ and kSZ power approaches zero for very negative values
of $\xi$.  
This is because the correlated power is proportional to the
tSZ power, and the tSZ power approaches zero in this limit.  

Decreasing tSZ power with increasing anti-correlation contradicts our naive expectations. 
In single-frequency bandpowers (e.g.~only 150\,GHz), increasing tSZ-CIB anti-correlations would increase the allowed tSZ contribution for any observed power level. 
However, this picture changes when we consider the other SPT frequency combinations. 
A tSZ-CIB anti-correlation reduces the power at $150\times150$ (which could be compensated by increasing tSZ power) but it also has $\sim$50\% larger effect at $150\times220$ (where there is effectively zero tSZ power). 
Conversely, a tSZ-CIB anti-correlation reduces the power at $95\times150$ by a similar amount to $150\times150$, whereas there is more tSZ power in the 95\,GHz band.  
Adding kSZ power more effectively cancels the tSZ-CIB correlation term than adding tSZ power in the most sensitive combinations of the three frequencies.

Due to this degeneracy, introducing a tSZ-CIB correlation substantially degrades the constraints on both the tSZ and kSZ power. 
In a model with $\xi$ as a free parameter, the measured tSZ power is $D_{3000}^{\rm tSZ}=3.26 \pm 1.06 \,\mu{\rm K}^2$. 
The 95\% confidence upper limit on the kSZ power is now $D_{3000}^{\rm kSZ} < 6.7 \,\mu{\rm K}^2$. 
The SPT data better constrain the linear combination of  $D_{3000}^{\rm tSZ} + 0.5 D_{3000}^{\rm kSZ} = 4.60 \pm 0.63 \,\mu{\rm K}^2$, although the limiting case with zero tSZ and high kSZ power is disfavored. 
The expanded parameter space resembles the constraint in previous SPT work (L10, S11); the additional frequency information in this analysis is constraining the CIB model instead of breaking the degeneracy between tSZ and kSZ power. 
To facilitate comparison we express the constraint in the same basis as S11; we find a constraint of $(1.034) D_{3000}^{\rm tSZ} + 0.5 D_{3000}^{\rm kSZ} = 4.71 \pm 0.64 \,\mu{\rm K}^2$. 
The multiplicative factor in parentheses before the tSZ term corrects for the different effective bandcenters in the two datasets. 
S11 found $D_{3000}^{\rm tSZ} + 0.5 D_{3000}^{\rm kSZ} = 4.5 \pm 1.0 \,\mu{\rm K}^2$.  
The two measurements are completely consistent with each other. 
We will discuss the inferred correlation and its impact on CIB constraints in \S\ref{sec:cibconstraint}.

\subsubsection{Beyond $\Lambda$CDM}
\label{sec:cmbextensions}

Lastly, we consider whether the derived SZ and CIB powers would change with an alternate cosmological model. 
Current measurements of the primary CMB anisotropy contain hints of parameters beyond the six $\Lambda$CDM parameters (\citealt{reichardt09a,komatsu11,dunkley11}; K11). 
The evidence for these model extensions is not yet compelling; however, it is interesting to test whether the extended models affect the SZ constraints. 
We consider three extensions: running of the scalar spectral index of primordial fluctuations, massive neutrinos, and extra neutrino (or other relativistic particle) species. 
We find no evidence for a degeneracy between any extended parameters and the kSZ or tSZ power. 
The kSZ and tSZ constraints are unchanged under these three $\Lambda$CDM model extensions. 

We also consider using an unlensed primary CMB power spectrum. 
We see a significant increase in the kSZ power when lensing is turned off. 
The inferred kSZ power increases by about the same amount as the primary CMB decreases in the absence of gravitational lensing at $\ell \sim 3000$. 
This produces a significant, $\sim 3\, \mu{\rm K}^2$, shift in the kSZ power. 
However, current CMB data rule out no-lensing scenarios at more than $5\,\sigma$ (K11). 
The induced uncertainty on the kSZ level for the current constraint on the lensing amplitude, $A_{lens}^{0.65}=0.95 \pm 0.15$, is only $0.4 \,\mu{\rm K}^2$.
This is much smaller than the $1\,\sigma$ uncertainty on kSZ, and therefore can be neglected.

 \section{Thermal SZ Interpretation}
\label{sec:tszconstraint}

We now focus on comparing our measurements of the tSZ 
power spectrum with the predictions of recent models and
simulations. 
We combine measurements of the primary CMB power spectrum, H$_0$, and BAO
with the tSZ signal to measure $\sigma_8$ and the sum of the neutrino masses, $\sum m_\nu$.

\subsection{Thermal SZ constraints}

\begin{figure*}[t]\centering
\includegraphics[width=0.9\textwidth]{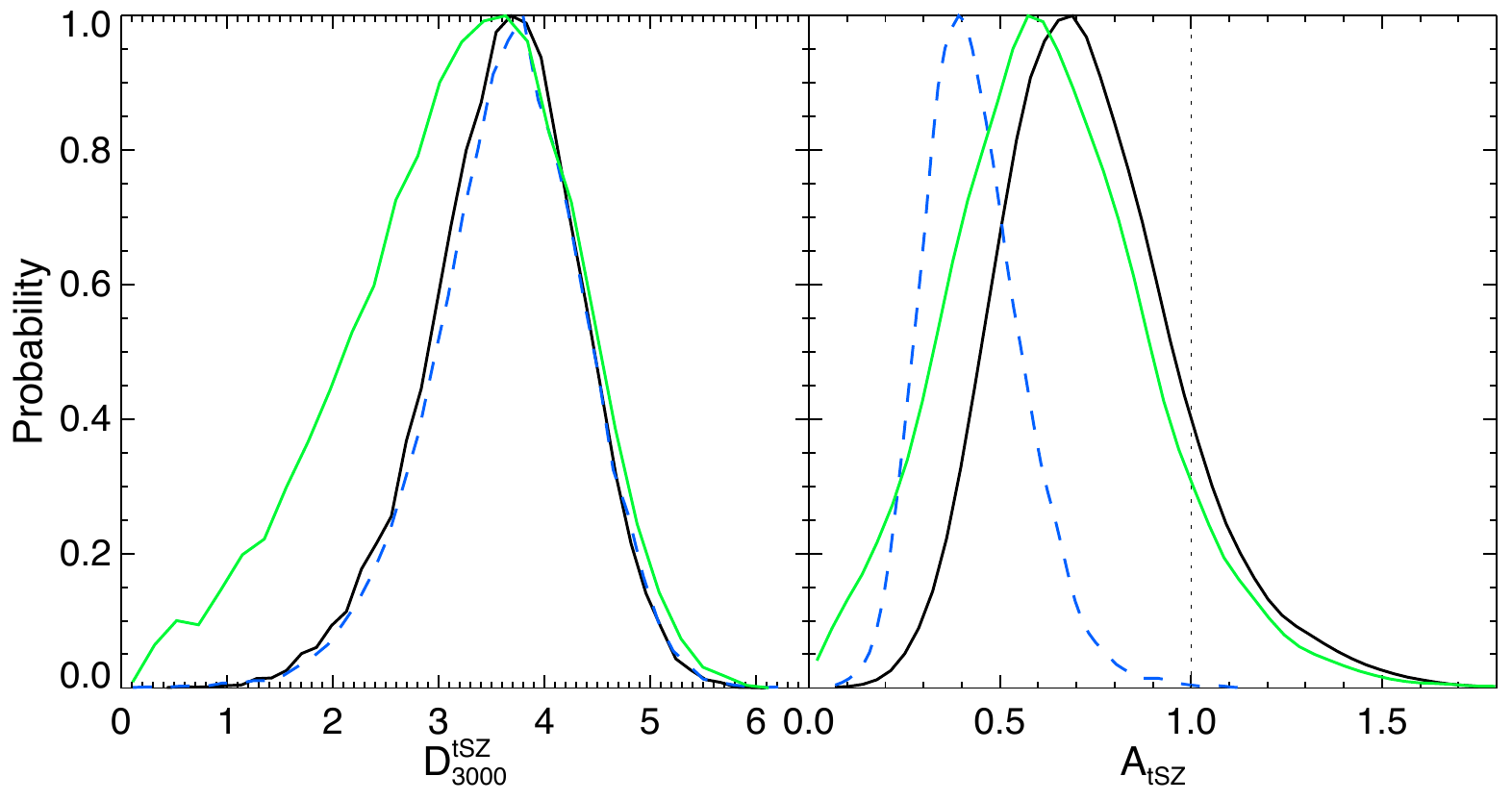}

  \caption[]{ 1D likelihood curves for the tSZ power spectra in $\mu {\rm K}^2$ at 150\,GHz and $\ell=3000$ (\textbf{\textit{Left panel}}) and ratio of measured to predicted tSZ power, $\atsz$ (\textbf{\textit{Right panel}}). 
    In the right panel, the vertical dotted line marks the expectation value for $\atsz$ ($\pm 0.04$ due to sample variance). 
  All curves assume  the CSF kSZ template, and modified BB CIB frequency scaling -- the tSZ results are unaffected by these choices. 
  The {\bf black solid line} shows the results with the Shaw tSZ model. 
  Results for the Sehgal tSZ model are shown by the {\bf blue dashed line}. 
  The power constraints (left) are identical, leading to much lower $\atsz$'s (right) as the Sehgal model predicts more tSZ power for a given cosmology. 
  Finally,  if we allow for possible correlations between the CIB and tSZ power spectra, the tSZ likelihood expands towards lower tSZ power (\textbf{green line}). 
  In the correlation case, we have assumed the Shaw tSZ model -- the impact is similar for the Sehgal model. 
  
}
  \label{fig:1dtsz}
\end{figure*}

The models for the tSZ power spectrum are based on simulations with a fixed cosmology. 
Therefore, to correctly compare the measured tSZ power with
theoretical predictions, we must rescale the theoretical power spectra so
that they are consistent with the cosmological parameters at each
point in the MCMC chains.  
Recall that the cosmological parameters are
derived solely from constraints on BAO, H$_0$, and the
primary CMB power spectrum and so are not directly influenced by the
amplitude of the measured tSZ power.

Following S11, we define a dimensionless tSZ scaling parameter,
$\atsz$, where
\begin{equation}
A_{\rm tSZ}(\kappa_i) = \frac{D_{3000}^{\rm tSZ}}{\Phi_{3000}^{\rm tSZ}(\kappa_i)} \;.
\label{eq:atszcosm}
\end{equation}
$D_{3000}^{\rm tSZ}$ is the measured tSZ power at $\ell = 3000$ and
$\Phi_{3000}^{\rm tSZ}(\kappa_i)$ is the template power spectrum
normalized to be consistent with the set of cosmological parameters
`$\kappa_i$' at step $i$ in the chain. We obtain the preferred value
of $\atsz$ by combining the measurement of $D_{3000}^{\rm tSZ}$ with the
cosmological parameter constraints provided by the primary CMB
measurements. As described in Section \ref{sec:tsz}, we consider two
template models for the tSZ power spectrum, the `Shaw' and `Sehgal' models. 
These two templates bracket the range of tSZ models considered by S11. 

To determine the cosmological scaling of
$\Phi_{3000}^{\rm tSZ}(\kappa_i)$, we run the Shaw model at each step in
the chain.  The approximate cosmological scaling of the Shaw model
around the fiducial model cosmological parameters is given by Equation
\ref{eq:tsz_scaling}.

In the right panel of Figure \ref{fig:1dtsz}, we show the 
constraints obtained on $\atsz$. The solid black line in each plot
shows the results for the baseline model using the Shaw tSZ template, the CSF kSZ
template, and the modified BB CIB frequency scaling, while the green line
shows those obtained when a free tSZ-CIB correlation is allowed in our
modeling.  The blue dashed line shows the results obtained using the
Sehgal tSZ model instead.  The numerical constraints are also given in
Table \ref{tab:asz_constraints}. Recall that a value of $\atsz \simeq
1$ implies that the measured tSZ power is consistent with the tSZ
template predictions, while $\atsz$ less (greater) than 1 indicates
that the model overestimates (underestimates) the amplitude.  Note
that sample variance will cause deviations from unity on the
order of 0.04.

The preferred value of $\atsz$ for the Shaw template is $1.2\,\sigma$
below unity (from the constraints given on $\ln{\atsz}$), while the
preferred value of $\atsz$ for the Sehgal template is $2.9\, \sigma$
below unity.  Although both templates predict more tSZ power than is
measured, the Sehgal template is in significant tension with the data.
As an aside, we note that using the Single SED CIB frequency scaling
instead of the modified BB has little effect on the tSZ amplitude.
Including a tSZ-CIB spatial correlation shifts the preferred value of
$\atsz$ to a slightly lower value, but principally increases the error
bars (by roughly $33\%$).

In Table \ref{tab:szchisq}, we evaluate the goodness of fit of our SZ
models by running new MCMC chains in which the tSZ and kSZ terms have been
fixed to their predicted, although cosmologically-rescaled, values.  All four
combinations of the Sehgal and Shaw tSZ models and the NR and CSF kSZ
models are considered. We show the change in the best-fit $\chi^2$ for
each case relative to our baseline model (Shaw tSZ, CSF kSZ).
Models that predict more SZ power are a slightly worse fit to the
data. Table \ref{tab:asz_constraints} shows that the measured tSZ
power is a factor of $0.41 \pm 0.13$ times that predicted by the Sehgal
template. This template predicts a tSZ amplitude that is poorly matched
to the data, resulting in a $\Delta \chi^2 = +8.4$ relative to the
baseline model. Likewise, the NR kSZ template predicts 1 \muksq\ more
power than the CSF kSZ template and thus provides a worse fit to the
measured powers, increasing $\chi^2$ by 2.1 (for Shaw tSZ) and $2.3$
(Sehgal tSZ).  
Adding a kSZ contribution from patchy reionization
would further increase the model tension.

The consistency between the SZ models and measured bandpowers can
be tested further by including additional external cosmological
constraints.  Both the tSZ and kSZ power spectra scale most
sensitively with $\sigma_8$.  We therefore importance sample each of
the MCMC chains presented in Table \ref{tab:szchisq} imposing a prior of
$\sigma_8 (\Omega_M/0.25)^{0.47} = 0.813 \pm 0.027$, based on the
X-ray cluster abundance measurements of \citet{vikhlinin09}. The
subsequent change in $\chi^2$ is given in the third column of the
table.  The prior from \citet{vikhlinin09} increases the significance
of the discrepancy with models that predict more tSZ power.  This
result is particularly interesting since it eliminates one possible
solution for the high-power tSZ models -- massive neutrinos.  The CMB
data alone is consistent with high neutrino masses which slow the growth of 
structure (see \S\ref{sec:mnu}).  If the sum of neutrino masses
was around 1 eV and $\sigma_8 \simeq 0.6$, the measured tSZ power
would in fact be higher than that predicted by the Sehgal model.  The
X-ray constraint, $\sigma_8 \simeq 0.8$, which is independent of the tSZ modeling,  rules out this line of argument.

\begin{table*}[ht!]
\begin{center}
\caption{\label{tab:asz_constraints} $\atsz$, $\aksz$ constraints}
\small
\begin{tabular}{c|c|c|c|c}
\hline\hline
\rule[-2mm]{0mm}{6mm}
Model & $\atsz$ & $\ln{\atsz}$ & $\aksz$ & $\sigma_8~ [\atsz \simeq 1]$\\
\hline
Shaw, CSF, modified BB & $0.70 \pm 0.21$ & $-0.37 \pm 0.30$ & $ <1.68$ & $0.797 \pm 0.011$\\
Sehgal, CSF, modified BB & $0.41 \pm 0.13 $ & $-0.89 \pm 0.31 $ & $ <1.57$ & $0.767 \pm 0.009$\\
Shaw, CSF, modified BB + tSZ-CIB  & $0.60 \pm 0.24$ & $-0.54 \pm 0.41 $ & $ < 3.98$ & $0.796 \pm 0.012$\\
\hline
\end{tabular}
\tablecomments{$\atsz$, $\aksz$, and $\sigma_8$ constraints. The distributions
  shown in Figure \ref{fig:1dtsz} are clearly non-Gaussian,
  therefore we give constraints on $\log{\atsz}$ as well. Constraints on
  $\aksz$ are $2\,\sigma$ upper limits. 
  Constraints on $\sigma_8$ include a prior on $\log{\atsz} = 0.00 \pm 0.04$; the width of this prior reflects sample variance but not model uncertainty. 
  The CMB+BAO+H$_0$ constraint on $\sigma_8$ is $0.812 \pm 0.018$. } \normalsize
\end{center}
\end{table*}

\begin{table}[ht!]
\begin{center}
\caption{\label{tab:szchisq} Delta $\chi^2$ for SZ models}
\small
\begin{tabular}{cc|cc}
\hline\hline
\rule[-2mm]{0mm}{6mm}
tSZ Model & homog. kSZ Model & \delchisq\ &\delchisq\ \\
&&& w. X-ray\\
\hline
Shaw & CSF & (0) & 0.0\\
Shaw & NR & 2.1 & 2.3 \\
Sehgal & CSF & 8.4 & 13.0\\
Sehgal & NR & 10.7 & 16.2\\
\hline
{\it Free} Shaw & {\it Free} CSF & -1.9 & -1.9\\
\hline
\end{tabular}
\tablecomments{ Change to the best-fit $\chi^2$ when the tSZ and kSZ
  contributions are fixed to the predictions of different models.
  Differences are reported relative to the Shaw tSZ / CSF kSZ case.
  The model predictions are scaled to account for the specific
  cosmology at each chain step as described in \S\ref{sec:tsz} and
  \S\ref{sec:hksz}.  Note that the last row has an additional two free
  parameters describing the amplitudes of the tSZ and kSZ power
  spectra.  } \normalsize
\end{center}
\end{table}

\subsection{$\sigma_8$ constraints}
\label{sec:sigma8}

To obtain joint constraints on $\sigma_8$ from both the primary CMB
anisotropy and the tSZ power spectrum, we must determine the probability
of observing the measured value of $\atsz$ for a given set of
cosmological parameters. In the absence of sample variance -- and
assuming the predicted tSZ template to be perfectly accurate -- the
constraint on $\sigma_8$ would be given by its distribution at
$\atsz=1$ for each of the models. In practice, the expected sample
variance of the tSZ signal is non-negligible. Based on the simulations
of \citet{shaw09}, L10 assumed the sample variance at $\ell = 3000$ to
be a lognormal distribution of mean $0$ and width $\sigma_{\ln
  (\atsz)} = 0.12$.  The survey area analyzed here is eight times that
of L10 and thus we reduce the width of this distribution by
$\sqrt{8}$. For each of the tSZ templates, we construct new MCMC chains by
importance sampling to include this prior.

\begin{figure}[t]\centering
\includegraphics[width=0.5\textwidth]{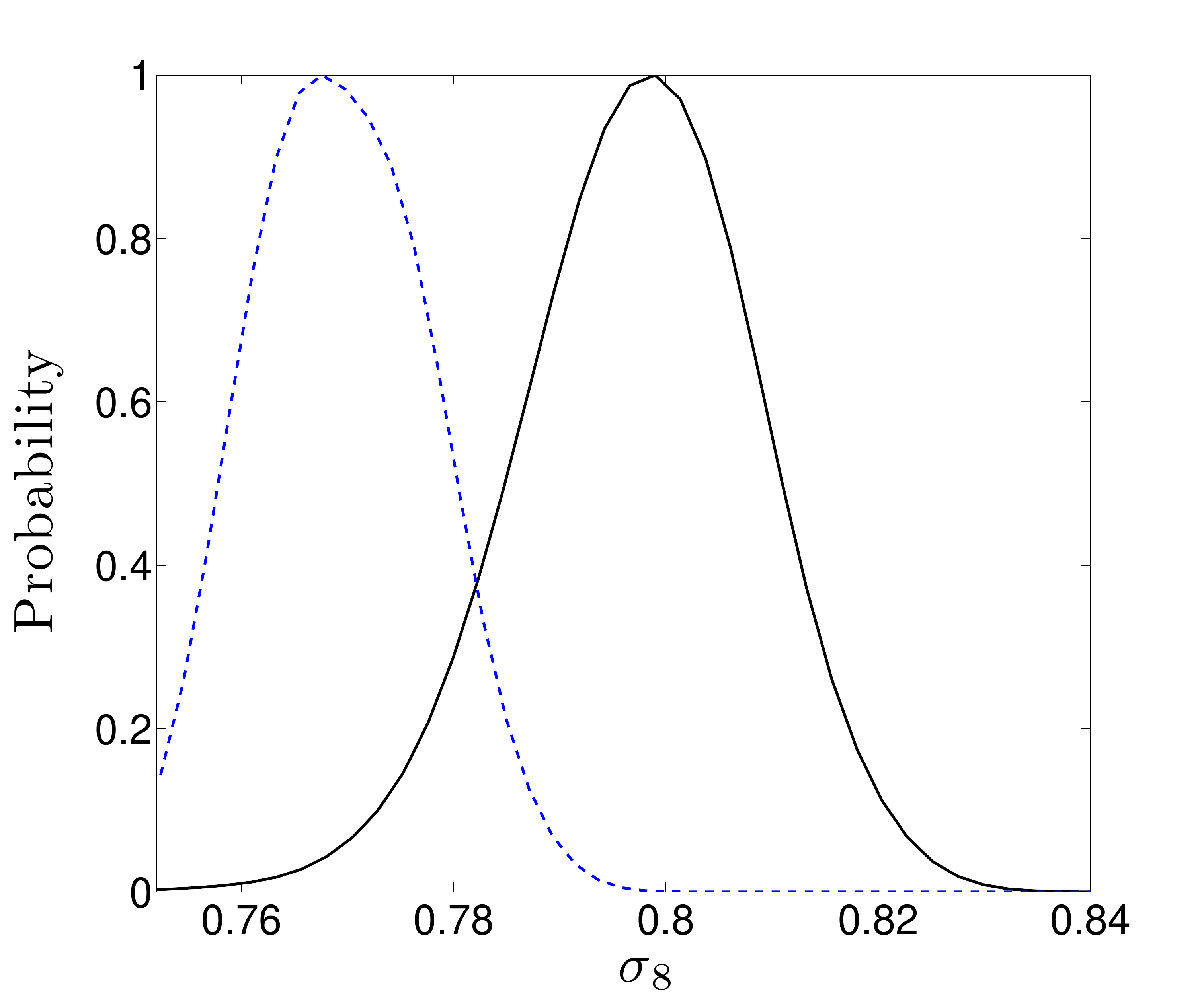}
\caption[]{$\sigma_8$ constraints for the Shaw ({\bf black solid line}) and Sehgal
  ({\bf blue dashed line}) tSZ templates. No model uncertainty has been
  included.}
\label{fig:sigma8}
\end{figure}

The 1D $\sigma_8$ constraints for the two tSZ templates are shown in
Figure \ref{fig:sigma8} and Table \ref{tab:asz_constraints}. Using the
Shaw template results in $\sigma_8 = 0.797\pm 0.011$, while the Sehgal
model prefers $\sigma_8 = 0.767 \pm 0.009$. The constraint without the
tSZ amplitude information (i.e. from the primary CMB + BAO + H$_0$
alone) is $\sigma_8 = 0.812 \pm 0.018$. Adding the tSZ information
therefore both reduces the size of the statistical error by 40\% and
lowers the preferred value, depending on the amplitude of the tSZ
template.

The difference in the preferred value of $\sigma_8$ provided by the
Shaw and Sehgal templates is significantly larger than the statistical
errors. This implies that the systematic
error due to the theoretical uncertainty in the tSZ modeling
significantly exceeds the statistical error. To produce a
constraint on $\sigma_8$ that accounts for this, we must add an
additional theory uncertainty to the sample variance when setting the prior on $\atsz$.

It is difficult to determine the level of theory uncertainty on the predicted
tSZ amplitude for a fixed set of cosmological parameters. A
significant fraction of the total tSZ power is contributed by groups
and clusters of mass $M_{500} < 2\times 10^{14} \hmsun$ and
redshift $z > 0.65$ \citep{trac11, battaglia11,shaw10}. To date,
such objects have not been studied in detail with SZ and X-ray
observations. From a theoretical perspective, models and simulations
have demonstrated that astrophysical processes such as feedback from
active galactic nuclei and turbulence-driven non-thermal pressure
support can significantly alter predicted cluster thermal pressure profiles, and
thus their SZ signal \citep{battaglia11b, trac11,parrish11,shaw10}.

Following S11, we choose to add a 50\% theory uncertainty on $\ln
(\atsz)$.  This uncertainty encompasses all four models considered by
S11.  As the sample variance is insignificant compared to the theory
uncertainty, the new prior on $\atsz$ is effectively a lognormal
distribution of width $\sigma(\ln{\atsz}) = 0.5$. With this assumption, 
we obtain
constraints of $\sigma_8 = 0.807 \pm 0.016$ for the Shaw model.  

We estimate the contribution of the systematic theory uncertainty to the total uncertainty quoted above as follows. 
We form a new prior based on the median value of  $\atsz = 0.7$ with the sample-variance based width of $\sigma(\ln
  (\atsz)) = 0.042$. 
  By sampling around the preferred value of $\atsz$ instead of unity, we avoid introducing artificial tension if the modeled amplitude is incorrect. 
  Importance sampling the MCMC chain with this prior leads to parameter constraints that include only statistical noise and sample variance. 
  The resulting $\sigma_8$ constraint is $\sigma(\sigma_8) = 0.010$. 
  Assuming the errors add in quadrature, this result implies a systematic modeling uncertainty of $\sigma(\sigma_8) = 0.012$. 
  
Given the current modeling uncertainties, the measured amplitude of tSZ
power in this work only marginally improves our knowledge of
$\sigma_8$ over existing constraints. A detailed and systematic study
of the thermal pressure profiles of low-mass ($\approx 10^{14} \hmsun$) clusters at $z \approx
0.65$ is required in order to reduce the theory uncertainty and thus
increase the precision of $\sigma_8$ constraints derived from the tSZ
power spectrum.

\subsection{$\sum m_\nu$ constraints}
\label{sec:mnu}

\begin{figure*}[t]\centering
\includegraphics[width=\textwidth]{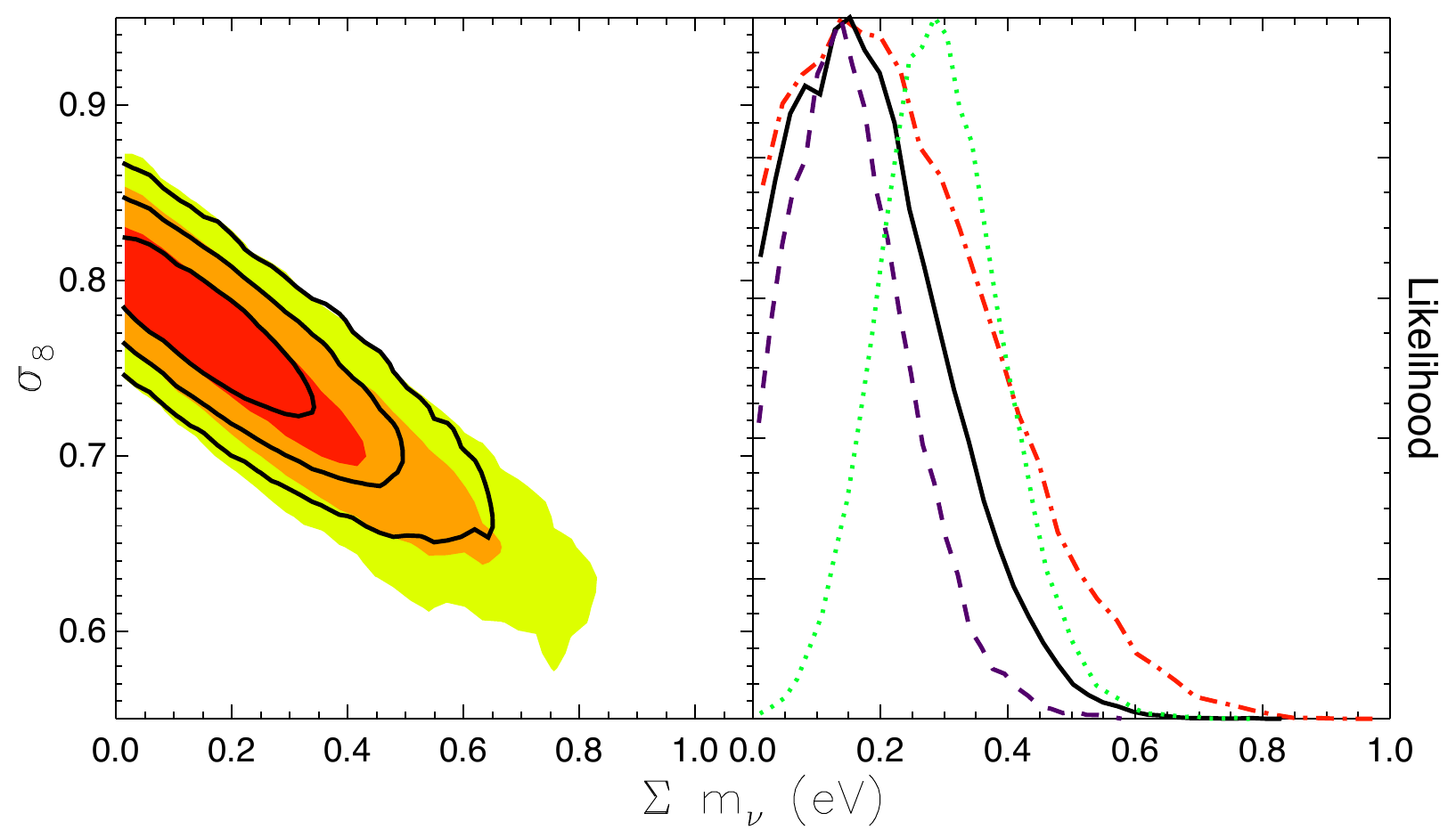}
\caption[]{ 
\textbf{\textit{Left panel:}} 2D likelihoods for $\sigma_8$ and $\sum m_\nu$. 
The {\bf filled colored contours} show the 1, 2, and 3\,$\sigma$ constraints for the CMB + H0 + BAO data. 
The {\bf  black solid lines} show the constraints with the addition of the tSZ information (assuming a 50\% theory prior). 
\textbf{\textit{Right panel:}} 1D likelihoods for the sum of neutrino masses for four cases. 
The {\bf red dot-dash line} shows the constraints without including tSZ data. 
The {\bf black solid  line} shows the constraints when we include the measured tSZ power with the Shaw tSZ model and a 50\% theory prior. 
This tightens the upper limit on the sum of the neutrino masses from 0.52\,eV to 0.40\,eV at 95\% CL. 
We show the constraints on neutrino mass that would be derived with {\it perfect} tSZ modeling in the final two lines: with the Shaw tSZ model ({\bf purple dashed line}) and Sehgal tSZ model ({\bf green dotted line}). 
The Sehgal model predicts more tSZ power for a given $\sigma_8$. 
As a result it pushes the allowed parameter volume to lower $\sigma_8$ and higher $\sum m_\nu$, leading to a nominal detection of the total neutrino mass at $0.29 \pm 0.10\,$eV.
 }
\label{fig:mnu}
\end{figure*}

We can apply the methodology outlined above to other cosmological parameters, most notably the sum of the neutrino masses, $\sum m_\nu$.
For CMB + H$_0$ + BAO, neutrino masses are highly degenerate with $\sigma_8$ as shown in Figure \ref{fig:mnu}. 
Higher neutrino masses lead to lower $\sigma_8$ since massive neutrinos slow the growth of structure below the neutrino free-streaming length. 
Introducing massive neutrinos weakens the CMB + H$_0$ + BAO constraint on $\sigma_8$ to $\sigma_8 = 0.756 \pm 0.044$. 
The sum of the neutrino mass is constrained to be less than 0.52\,eV at 95\% CL. 

The tSZ power spectrum presents an independent probe of $\sigma_8$
that breaks the $\sum m_\nu$ degeneracy, and thereby improves the
neutrino mass determination.  The preference for positive neutrino
masses increases with models that predict more tSZ power.  Models that
predict more tSZ power require a lower $\sigma_8$ to match the
observed tSZ power.  As can be seen in Figure \ref{fig:mnu}, lower
$\sigma_8$ values favor larger neutrino masses.

With massive neutrinos and no modeling uncertainty (but including a 4\% sample variance),
adding the tSZ information reduces the uncertainty on $\sigma_8$ by a
factor of 2 - 3.  The median $\sigma_8$ moves up slightly to $\sigma_8
= 0.776 \pm 0.019$ with the Shaw model and down slightly to $\sigma_8
= 0.732 \pm 0.017$ with the Sehgal model.  In this optimistic
scenario, we would find $\sum m_\nu = 0.15 \pm 0.09\,$ or $0.29 \pm
0.10\,$eV for the Shaw and Sehgal models, respectively.  
Note that the X-ray $\sigma_8$ measurement mentioned earlier is in tension with the preferred $\sigma_8$ with the Sehgal model. 
With a more
realistic 50\% modeling uncertainty on the Shaw model, the constraint
is $\sigma_8 = 0.768 \pm 0.031$.  With this modeling uncertainty, the
Sehgal and Shaw model results are consistent at $\sim0.5\,\sigma$.
The upper limit on $\sum m_\nu$ is reduced from 0.52\,eV to 0.40\,eV
at 95\% CL with the addition of tSZ information.  Even with a large
theory prior, the SPT tSZ measurements improve the neutrino mass
constraints by disfavoring regions of parameter space with very low
$\sigma_8$ values.

We consider how the uncertainties on $\sigma_8$ and $\sum m_\nu$ break down between systematic and statistical uncertainties  using the  method presented in \S\ref{sec:sigma8}. 
The prior is constructed as before, and centered the preferred value of $\atsz=1.14$ in the $\sum m_\nu > 0$\,eV chain. 
 We find that the uncertainty on $\sigma_8$ can be allocated as $\sigma_{\rm sys}(\sigma_8) = 0.024$ (for the modeling uncertainty) and $\sigma_{\rm stat}(\sigma_8) = 0.020$ (for statistical noise plus sample variance). 
Similarly,   the uncertainty on $\sum m_\nu$ can be divided into $\sigma_{\rm sys}(\sum m_\nu) = 0.10$\,eV and $\sigma_{\rm stat}(\sum m_\nu) = 0.10$\,eV. 
The 50\% theory uncertainty is dominant for both parameters.

\section{kSZ interpretation}
\label{sec:kszconstraint}

\begin{figure*}[t]\centering
\includegraphics[width=0.9\textwidth]{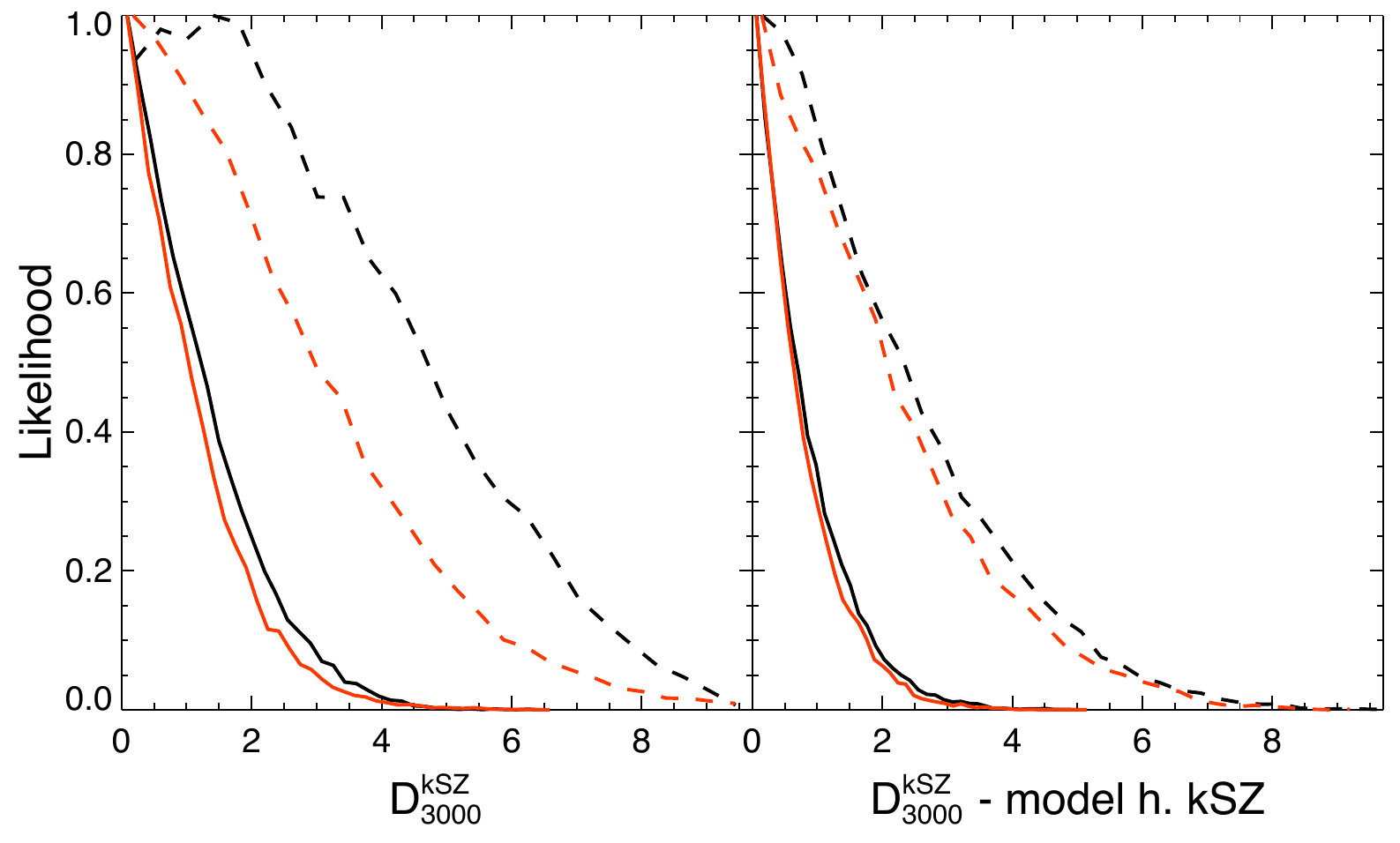}
  \caption[]{
1D likelihood curves for the kSZ power (in $\mu {\rm K}^2$). 
  \textbf{ \textit{ Left panel:}} Constraints on the kSZ power with different CIB models and kSZ templates. 
The modified BB CIB frequency scaling and Shaw tSZ template are assumed in all cases. 
 {\bf Solid lines} denote constraints without tSZ-CIB correlations; {\bf dashed lines} mark constraints with tSZ-CIB correlations allowed.
   {\bf Black lines} show the results with the CSF kSZ template (the NR kSZ template is similar).
    {\bf Red lines} mark the constraints when we alternatively use a template based on the kSZ contribution from the epoch of reionization. 
  \textbf{ \textit{  Right panel:}} The constraints on the patchy kSZ power when we include a model prediction for the homogeneous kSZ power. 
 {\bf Solid lines} denote constraints without tSZ-CIB correlations; {\bf dashed lines} mark constraints with tSZ-CIB correlations allowed.
 {\bf Black lines} show the results when the CSF homogeneous kSZ model is assumed. 
 This model includes cooling and star formation which suppresses the kSZ power. 
 {\bf Red lines} mark the constraints when the higher amplitude NR homogeneous kSZ model is assumed. 
 }
  \label{fig:Dksz}
\end{figure*}

\begin{figure}[thb]\centering
\includegraphics[width=.45\textwidth]{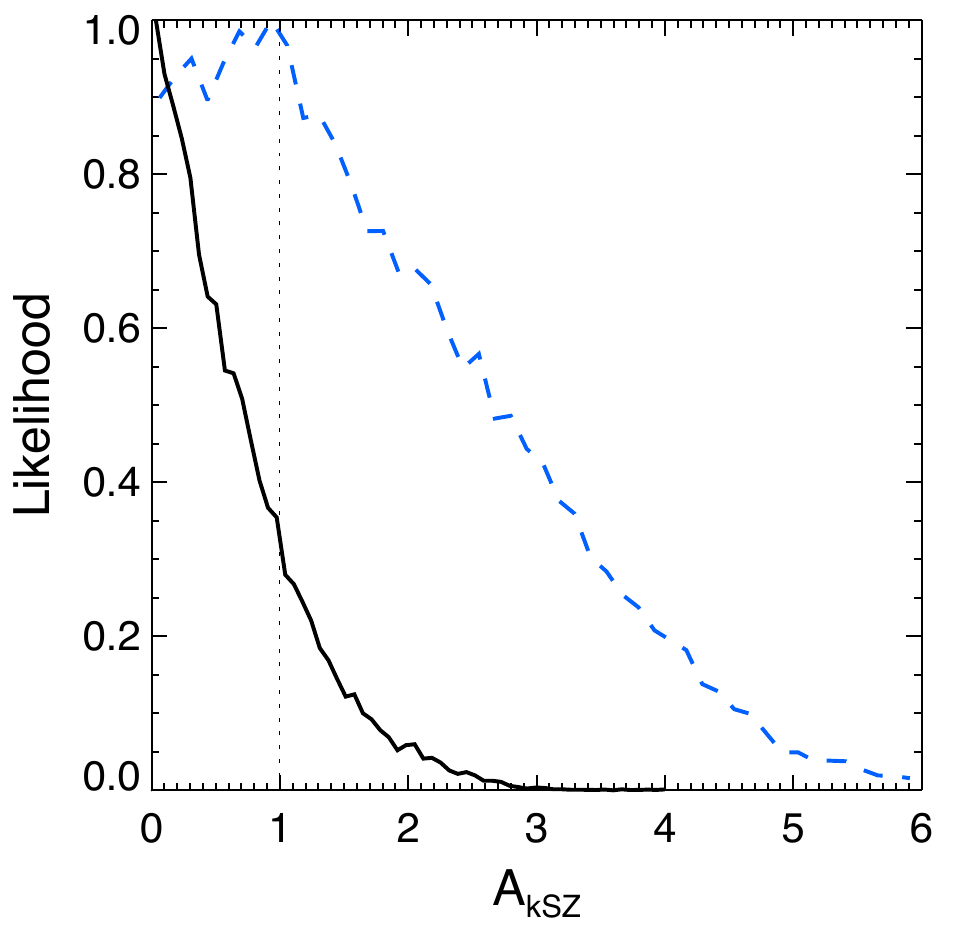}
\caption[]{Posterior distribution of the kSZ scaling parameter $\aksz$ (see
  Equation \ref{eq:atszcosm} and Table \ref{tab:asz_constraints}) for
  different  CIB models. 
  The vertical dotted line marks the model expectation value for $\aksz$ ($\pm 0.01$ due to sample variance). 
  The CSF kSZ model we consider here is the lower of the two kSZ models and does not include the kSZ contribution from patchy reionization. 
  Therefore, $\aksz > 1$ would be reasonable. 
The {\bf solid black line} shows
  the baseline model using the Shaw tSZ template, the CSF kSZ template,
  and the modified BB CIB frequency scaling. 
  The {\bf blue dashed line} shows
  the results obtained when the tSZ-CIB correlation is allowed in our
  modeling. }
\label{fig:Aksz_constraints}
\end{figure}

We next turn to comparing our measurements of the kSZ
power spectrum with the predictions of recent models and
simulations. 
We first compare our constraints on the amplitude of the kSZ power
spectrum with our homogeneous kSZ templates. 
We then briefly mention the implications of the kSZ measurement for the epoch of reionization. 
Z11 discuss what we can learn about the epoch of reionization in more detail. 

As previously presented, in the baseline model, we find the kSZ power
to be robust to template shape between the CSF, NR, and patchy kSZ
templates.  In contrast to the tSZ, we see a small dependence on the
frequency scaling of the CIB with slightly lower kSZ upper limits for
the Single SED scaling than for the baseline modified BB model.  However,
the major modeling uncertainty, as with the tSZ, is in the tSZ-CIB
correlation (see Figure  \ref{fig:Dksz}).  This correlation is highly
degenerate with the kSZ power (see Figure  \ref{fig:2dkszcorrel}) and,
when introduced as a free parameter, nearly triples the allowed kSZ
95\% upper limit.  
The kSZ template shape affects the kSZ upper limit at the $\sim 20$\% level
when tSZ-CIB correlations are allowed.

\subsection{Model comparison}

In an identical manner to the tSZ, we define a dimensionless scaling
parameter $\aksz$ for the homogeneous kSZ power. We use the NR and CSF
kSZ models of \citet{shaw11}.  Equation \ref{eq:ksz_scaling} gives the
scaling of the CSF kSZ template $\Phi_{3000}^{\rm kSZ}$ with cosmological
parameters.  

In Figure \ref{fig:Aksz_constraints}, we show the  constraints
obtained on $\aksz$. The solid black
line shows our baseline model using the Shaw tSZ
template, the CSF kSZ template and the modified BB CIB frequency
scaling.  
The blue dashed  line shows those obtained when a free tSZ-CIB correlation is
included in our modeling. The numerical constraints are also given in
Table \ref{tab:asz_constraints}. Recall that a value of $\aksz = 1$
implies that the measured kSZ power is consistent with the kSZ
template predictions, while $\atsz$ less (greater) than 1 indicates
that the model overestimates (underestimates) the  amplitude.  Note
that sample variance will cause deviations away from unity on the
order of 0.01.

In this work, we have significantly tightened the upper limits on the
kinetic SZ power.  These limits are not in significant tension
with our fiducial homogeneous kSZ template \citep[the CSF model
  of][]{shaw11}.  The $2\sigma$ upper limit on $\aksz$ exceeds 1 for
all of the MCMC chains in Table \ref{tab:asz_constraints}. 
  However, for our baseline model,
the upper limit is only modestly in excess of 1.  
The Single SED CIB model allows for slightly less kSZ power than the modified BB frequency scaling.  Therefore, while the data are consistent
with the CSF kSZ prediction, there is little room for an additional
contribution from patchy reionization.  Z11 note that, if the duration of
reionization spans $\Delta z = 2$ we expect an additional 1\,\muksq\ of kSZ
power. For $\Delta z = 4$ we expect 2\,\muksq. Clearly when combined
with the post-reionization signal, the predicted kSZ will
significantly exceed that measured by SPT in the baseline model. 

However, as discussed in \S\ref{sec:tszcibbaseresults}, incorporating
a tSZ-CIB spatial correlation into our model allows for considerably
more kSZ power than our fiducial model. In this case, the upper
limit on $\aksz$ is 4.13, which allows for substantially more power
than predicted by the homogeneous kSZ models alone, and thus a
significant contribution from patchy reionization.

\subsection{Patchy reionization}

We now switch to considering limits on the kSZ signal contributed by
patchy reionization.  For this we use a patchy kSZ template from Z11
based on a model in which reionization started at $z=11$ and ended $z=8$.  As
would be expected, the kSZ signals from the reionization and 
post-reionization era are nearly perfectly degenerate.  Therefore, our
constraints on the patchy kSZ power depend on the modeling of the
post-reionization homogeneous kSZ power.  We choose to bracket the
latter with three cases: no homogeneous kSZ, the intermediate CSF
homogeneous kSZ model, and the highest NR homogeneous kSZ model.

The constraints on patchy kSZ power are shown in Figure \ref{fig:Dksz}.
Note that we see less difference in the allowed patchy kSZ power than would be inferred from the power differences between the assumed homogeneous kSZ model. 
The smaller than expected differences are because the likelihood function for the total kSZ power, before any amount of homogeneous kSZ power is subtracted, already peaks at zero. 
We find 95\% CL upper limits on the kSZ power due to patchy reionization of 2.6, 2.0, and $1.9\,\mu{\rm K}^2$ for the zero, CSF, and NR homogeneous kSZ models respectively. 
As with all the kSZ constraints, we find the patchy kSZ constraints are most sensitive to tSZ-CIB correlations.
With a free tSZ-CIB correlation and the same set of three homogeneous kSZ models, the 95\% CL upper limits on patchy kSZ power are degraded to 5.7, 4.9, and $4.7\,\mu{\rm K}^2$. 
Z11 interpret these upper limits on patchy kSZ power to constrain the duration of the epoch of reionization and the ionization history of the Universe.

\section{Cosmic infrared background constraints}
\label{sec:cibconstraint}

\begin{figure*}[t]\centering
\includegraphics[width=0.9\textwidth]{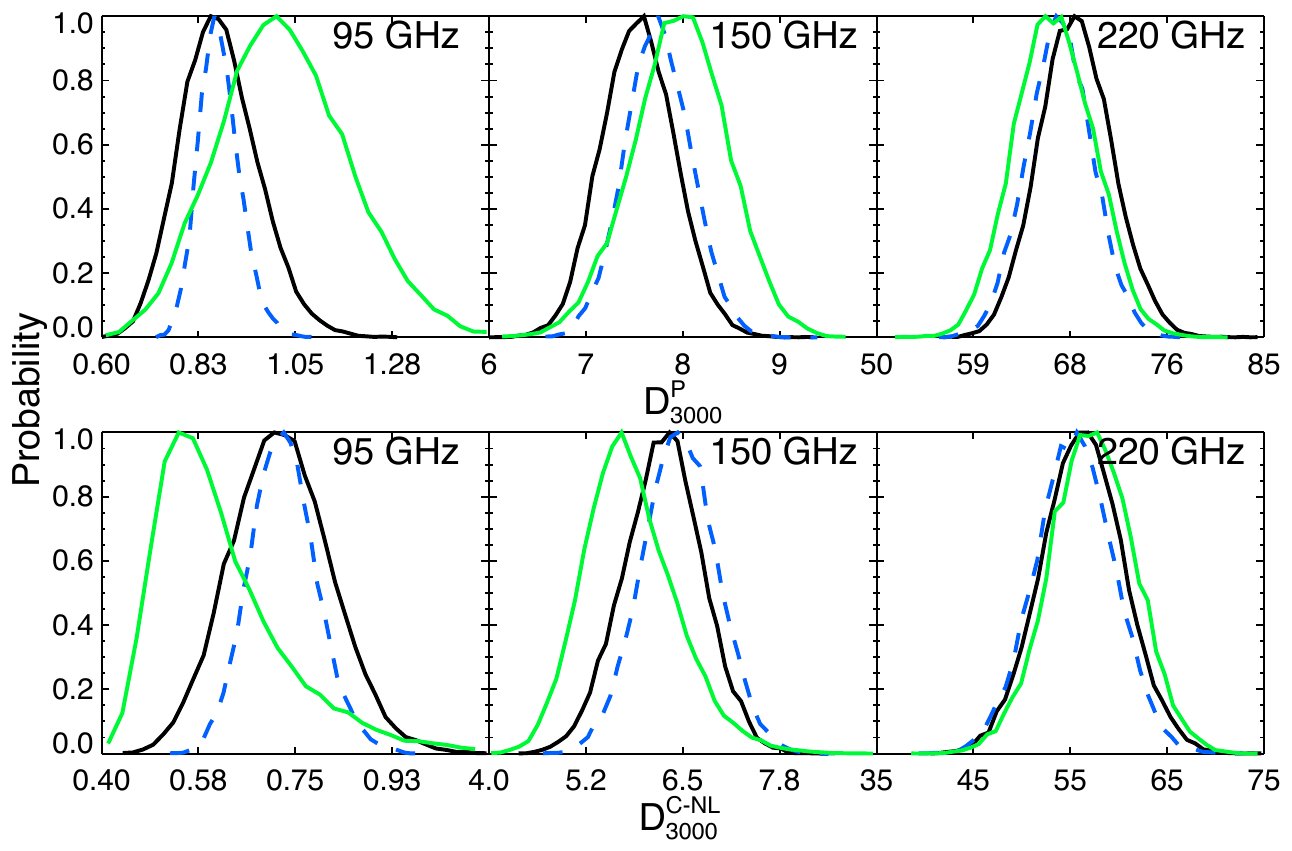}
  \caption[]{ 1D likelihood functions for the powers in $\mu{\rm K}^2$ of the two CIB components at $\ell = 3000$ in the three SPT bands. 
  The top row shows the constraints on Poisson power; the bottom row shows constraints on the power-law clustering power. 
  Sources detected with flux $>6.4\,$mJy at 150\,GHz are not included. 
  From left to right, the columns display the constraints at 95, 150, and 220\,GHz. 
  The likelihood functions with the fiducial, modified BB CIB frequency scaling are shown by the {\bf black lines}. 
The {\bf green lines} display the likelihood functions if the clustered and Poisson DSFG components are allowed to scale independently with frequency.
If we use the Single SED CIB frequency scaling instead, we find the {\bf blue dashed lines}.
The CIB constraints for the Single SED and modified BB scalings are very similar. 
The CIB powers, particularly the extrapolation to $95\,$GHz,  change somewhat when the Poisson and clustered component frequency scalings vary independently. 
The preferred spectral index is slightly shallower for the Poisson than clustered component. 
The Poisson and power-law clustering terms are comparable on angular scales near $\ell = 2500$. 
  }
  \label{fig:cibconstraints}
\end{figure*}

The combination of \planck\ and SPT data strongly constrains the CIB power spectrum at millimeter wavelengths, especially at 220\,GHz. 
The CIB constraints for a variety of model assumptions are summarized in Table \ref{tab:cibparams} and Figure \ref{fig:cibconstraints}. 
Recall that, in all cases, sources with flux $>$6.4\,mJy at 150\,GHz have been masked. 
We find the 220\,GHz Poisson amplitude at  $\ell = 3000$ to be $D^{p}_{3000}=68.0 \pm 3.3\, \mu {\rm K}^2$. 
The clustering amplitude is $D^{c}_{3000}=56.3 \pm 4.2\, \mu {\rm K}^2$ -- about 45\% of the total CIB power at this angular scale. 
The cross-over point where the  Poisson and clustering terms have equal power is at $\ell = 2500$. 
At 150\,GHz, the Poisson and clustering powers are $D^{p}_{3000}=7.54 \pm 0.38\, \mu {\rm K}^2$ and $D^{c}_{3000}=6.25 \pm 0.52\, \mu {\rm K}^2$ respectively. 
The estimated CIB  at 95\,GHz is very small, but this is an extrapolation from higher frequencies and sensitive to the frequency modeling. 
Both Poisson and clustered CIB components have more power at 150\,GHz than the SZ spectra across the multipoles ($\ell > 2000$) measured in this work. 
We will leverage the wide range of angular scales probed by SPT and \planck\ to constrain the shape of the CIB clustering in \S\ref{subsec:powerlaw}. 
 
Both models considered for the frequency scaling of the CIB lead to similar results. 
The data require a steep drop-off in CIB power from 220 to 150\,GHz; the spectral index is $\alpha_{150/220} = 3.51 \pm 0.04$ and $\alpha_{150/220}  = 3.56 \pm 0.07$ for the Single SED and modified BB models respectively. 
This is shallower by $1\,\sigma$ than the $\alpha_{150/220}  = 3.68 \pm 0.07$ measured for the clustered spectral index by \citet{addison11} using data from  BLAST, \planck, and ACT. 
We note that when we allow the Poisson and clustered spectral indices to differ later in this section, the inferred clustered spectral index becomes more consistent with the \citet{addison11} result. 
The observed spectral index for the lowest two published \planck\ channels (217 and 353\,GHz) is shallower: $\alpha_{217/353} = 3.22\pm 0.08$. 

These indices are steeper than many current models predict. 
For instance, the best-fit parameters found by \citet{fixsen98} when fitting {\it COBE}--FIRAS data predict $\alpha_{150/220}  = 2.6$.
The \citet{planck11-6.6_arxiv} previously pointed out that these parameters also overpredict the CIB anisotropy from 217 - 857\,GHz. 
The alternate parameter fitting by \citet{gispert00} that works well across the \planck\ frequencies \citep{planck11-6.6_arxiv}  still predicts a shallower $\alpha_{150/220}  =3.04$ between 150 and 220\,GHz. 
\citet{bethermin11} modeled DSFG source counts using data from {\it IRAS}, {\it Spitzer}, {\it Herschel}, and AzTEC -- these modeled source counts were used to predict the Poisson power in the \planck\ CIB bandpowers \citep{planck11-6.6_arxiv}. 
The spectral index predicted by the B{\'e}thermin model is $\alpha_{150/220}  = 2.8$. 
The maximum spectral index among the ten provided realizations of the B{\'e}thermin model is $\alpha_{150/220} = 3.1$. 
These models are a poor fit to the observed CIB power at millimeter wavelengths. 
For instance, the B{\'e}thermin model predicts $\sim$25\% too much Poisson power at 220\,GHz and twice the observed power at 150\,GHz.

\begin{figure*}[tbh]\centering
\includegraphics[width=0.9\textwidth]{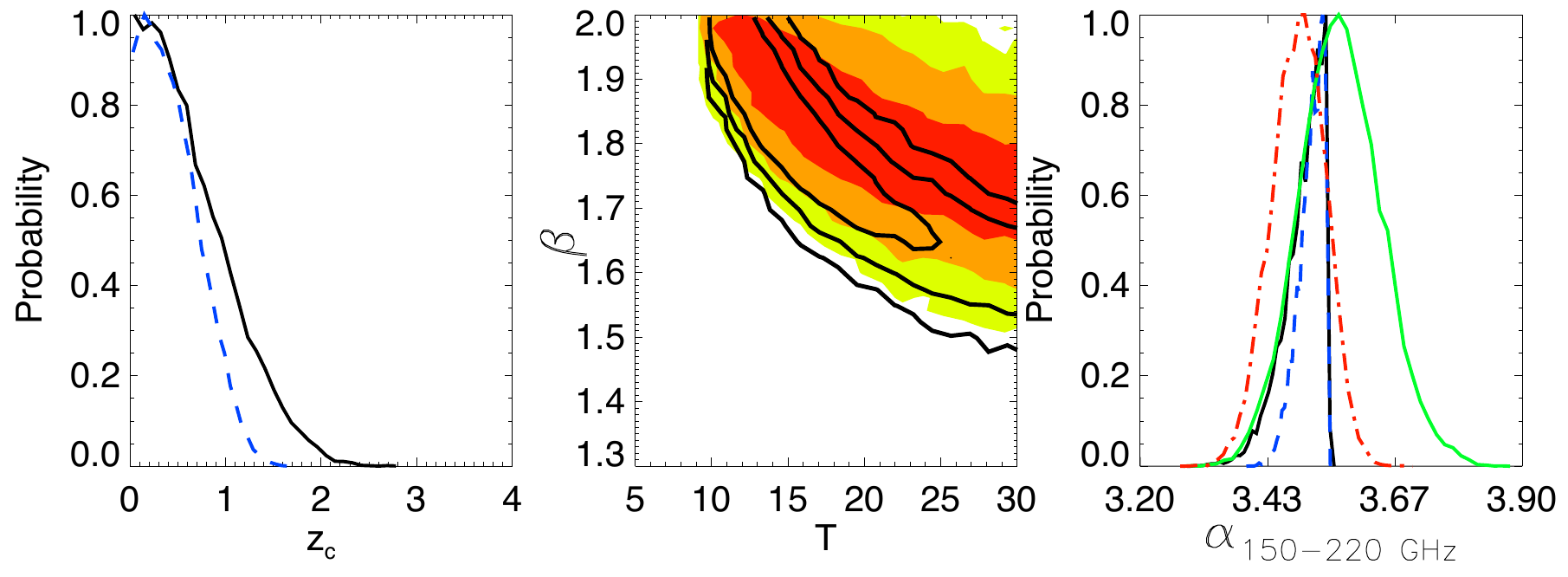}
  \caption[]{ CIB frequency modeling. 
  \textit{\textbf{Left panel:}} Constraint on the mean source redshift with the simple Single SED model parametrized with T = 35 K, $\beta = 2$, and $\sigma_z = 2$.
  The redshift constraints with and without the HFI 353 GHz data included are plotted in {\bf blue dashed} and  {\bf black lines}  respectively. 
    \textit{\textbf{Middle panel:}}
  2D likelihood surface for T and $\beta$ for the modified black-body CIB spectrum with 95-220\,GHz data ({\bf shaded contours}) and 95-353\,GHz data (\textbf{line contours}). 
  \textit{\textbf{Right panel:}}
  The effective CIB spectral index for 150 - 220\,GHz. 
  For the Single SED model frequency scaling, the preferred spectral indices are plotted in {\bf black solid} and {\bf blue  dashed lines} for 95-220\,GHz and 95-353~GHz respectively. 
    The sharp cutoff near $\alpha = 3.55$ is the \zcenter\ $> 0$ prior in the Single SED model. 
  The spectral index constraints with the modified black-body model with and without 353~GHz are shown by the {\bf red dot-dashed} and  {\bf green solid lines} respectively. 
}
  \label{fig:cibfreqconstraints}
\end{figure*}

There is a clear degeneracy between the two parameters of the modified BB model with only $\le 220\,$GHz data. 
Recall that in this model, the CIB spectrum is described by $\nu^\beta B_\nu(T)$. 
Without  353 GHz data, temperatures above $\sim$\,$15\,$K are constrained only by the prior since the observed frequencies are in the RJ tail of the spectrum. 
The preferred temperature is $>\,$12 K, and $\beta = 1.8 \pm 0.1$. 
These results are plotted in Figure \ref{fig:cibfreqconstraints}. 
Adding 353 GHz data partly breaks the T-$\beta$ degeneracy. 
We showed earlier in \S\ref{subsec:freqdep} that the spectral index was shallower in the higher frequency \planck\ data - $\alpha_{217/353} = 3.22\pm 0.08$. 
The changing slope means that the modified BB model prefers lower temperatures ($16 \pm 6\,$K, with a non-Gaussian distribution) and high values of $\beta$ ($\beta = 1.80 \pm 0.14$) when we include the 353 GHz data. 
The modified BB parameters, with and without 353\,GHz data, are consistent at $<\,1\,\sigma$.

With the Single SED frequency scaling, the median $\zcenter = 0.53$, but the best-fit $\zcenter \sim 0$. 
Recall that in the Single SED model, the DSFG density as a function of redshift is described by a Gaussian centered at \zcenter. 
The distribution of \zcenter\ tightens around zero when the 353\,GHz data are added. 
This suggests that either the adopted SED template is not representative or that using a single SED for all galaxies is an over-simplification. 
Note that in the Single SED model, the maximum allowed spectral index is $\alpha=3.57$ at \zcenter\ = 0. 
For the specified model parameters, higher spectral indices would require $\zcenter < 0$ which is not allowed by the \zcenter\ prior. 
The data are exploring the area at this limit, which slightly lowers the median spectral index  for the Single SED model. 
These redshifts are driven low to accommodate the steep observed spectral index. 
These constraints are shown in Figure \ref{fig:cibfreqconstraints}. 
The tighter spectral index range with the Single SED model also  (artificially)  tightens the kSZ constraints by a small amount.

We explore scaling the clustered and Poisson CIB independently with frequency, and find a small preference for different spectral indices when considering only $\le$\,220\,GHz data. 
For instance, with the modified BB model, the 150 - 220\,GHz spectral index is $\alpha_{150/220} = 3.45 \pm 0.11$ and $3.72 \pm 0.12$ for the Poisson and clustered term respectively -- a difference of $1.7\,\sigma$. 
The spectral index constraint is modestly weakened by the additional free parameter, but both parameters remain well-constrained. 
The SZ constraints are not affected. 
However, we note that adding the 353\,GHz \planck\ data removes this small preference for different spectral indices. 
In this case, the Poisson and clustered spectral indices are $\alpha_{150/220} = 3.46 \pm 0.09$ and $3.50 \pm 0.08$ respectively.

\begin{table*}[ht!]
\begin{center}
\caption{\label{tab:cibparams} CIB Constraints}
\small
\begin{tabular}{l|ccc|ccc}
\hline\hline
\rule[-2mm]{0mm}{6mm}
Model &  \multicolumn{3}{c}{Poisson ($\mu {\rm K}^2$)} & $\alpha_p^{150-220}$ & $\alpha_c^{150-220}$ & tSZ-CIB correlation\\
& 95\,GHz & 150\,GHz & 220\,GHz & &&\\
\hline
Mod. BB   & $ 0.87 \pm 0.09$ & $ 7.54 \pm 0.38$ & $67.96 \pm 3.29$ & $ 3.56 \pm 0.07$ & - & - \\  
~~$\alpha_c \ne \alpha_p$   & $ 1.02 \pm 0.15$ & $ 7.97 \pm 0.49$ & $65.99 \pm 3.65$ & $ 3.45 \pm 0.11$ & $ 3.72 \pm 0.12$ & - \\  
~~Linear th. clustering   & $ 0.90 \pm 0.10$ & $ 7.73 \pm 0.48$ & $69.10 \pm 3.88$ & $ 3.55 \pm 0.08$ & -& - \\  
~~Power law   & $ 0.86 \pm 0.10$ & $ 7.50 \pm 0.52$ & $67.79 \pm 4.23$ & $ 3.57 \pm 0.08$ & - & - \\  
~~tSZ-CIB corr.   & $ 1.04 \pm 0.15$ & $ 8.05 \pm 0.48$ & $66.67 \pm 3.68$ & $ 3.45 \pm 0.11$ & - & $-0.18 \pm 0.11$ \\  
Single Sed   & $ 0.87 \pm 0.05$ & $ 7.74 \pm 0.34$ & $66.51 \pm 2.96$ & $ 3.51 \pm 0.04$ & -& - \\  
\hline\hline
 & \multicolumn{3}{c}{Non-linear Clustering ($\mu {\rm K}^2$)}& \multicolumn{3}{c}{Lin. Theory Clustering ($\mu {\rm K}^2$)} \\
& 95\,GHz & 150\,GHz & 220\,GHz & 95\,GHz & 150\,GHz & 220\,GHz  \\
\hline
Mod. BB   & $ 0.72 \pm 0.09$ & $ 6.25 \pm 0.52$ & $56.26 \pm 4.22$ &  - & - & - \\  
~~$\alpha_c \ne \alpha_p$   & $ 0.59 \pm 0.11$ & $ 5.77 \pm 0.59$ & $57.26 \pm 4.34$ &  - & - & - \\  
~~Linear th. clustering   & $ 0.67 \pm 0.13$ & $ 5.73 \pm 1.00$ & $51.13 \pm 9.12$ &  $ 0.03 \pm 0.05$ & $ 0.26 \pm 0.43$ & $ 2.39 \pm 3.78$ \\  
~~Power law   & $ 0.72 \pm 0.11$ & $ 6.28 \pm 0.71$ & $56.70 \pm 6.39$ &  - & - & - \\  
~~tSZ-CIB corr.   & $ 0.87 \pm 0.17$ & $ 6.73 \pm 0.73$ & $55.85 \pm 4.47$ &  - & - & - \\  
Single Sed   & $ 0.73 \pm 0.06$ & $ 6.46 \pm 0.48$ & $55.48 \pm 4.10$ &  - & - & - \\

\hline
\end{tabular}
\tablecomments{  Constraints on the CIB power and spectral index are presented for six CIB models. 
Five of these models use the modified BB frequency dependence for the CIB. 
The first row is for the baseline model. 
The next four rows consider 1-parameter extensions to the baseline CIB treatment. 
In the ``$\alpha_c \ne \alpha_p$" row, the Poisson and clustered terms have independent frequency scalings. 
In the ``Linear th.~clustering" row, we allow for a (positive or negative) linear theory clustering term in addition to the power law clustering term. 
In the ``Power law" row, we introduce the exponent of the power law clustering term as a free parameter instead of fixing it to 0.8. 
In the ``tSZ-CIB corr." row, we allow for correlations between the tSZ and CIB components. 
The last row considers constraints under the alternate Single SED frequency modeling for the CIB.   
For each model, we present the constraints on the amplitudes in $\mu{\rm K}^2$ at $\ell = 3000$ of the Poisson and power-law clustering CIB templates. 
For the linear-theory clustering row, we also present the preferred amplitudes of the linear-theory term -- none of which represent detections. 
Finally, we present the effective spectral index from 150 to 220\,GHz for the Poisson (and if different) the clustering terms. 
  }
\normalsize
\end{center}
\end{table*}

\subsection{Angular dependence of the clustered CIB component}
\label{subsec:powerlaw}

As noted earlier, we find strong evidence for clustered DSFGs in the data. 
We have modeled the clustering by a power law with ${\it D}_\ell \propto \ell^{0.8}$; here we examine if the data prefer an alternate shape. 
The SPT data alone can be fit with either a linear-theory clustering template or the power-law baseline template -- only the extremely wide angular range probed by the combination of SPT and HFI allows  shape discrimination in the clustered template.
We parametrize the CIB clustering shape in two ways: fitting for the exponent of the power law, or fitting for the amplitude of a linear-theory term in addition to the power law term (with fixed exponent). 
Adding linear theory power is similar in effect to lowering the power law exponent.

\begin{figure}[tbh]\centering
\includegraphics[width=0.45\textwidth]{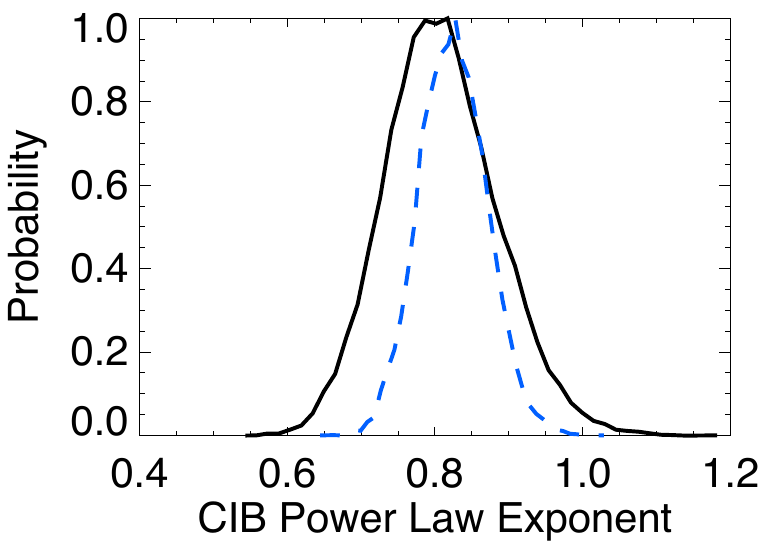}
  \caption[]{
  1D likelihood function for the power-law exponent, $\gamma$, of the CIB clustering term (${\sc D}_\ell \propto \ell^{\gamma}$) . 
  The fiducial model value is 0.8. 
  The curves for constraints with and without including the HFI 353~GHz CIB bandpowers are plotted 
with {\bf dashed blue} and {\bf black lines} respectively. 
   }
   
  \label{fig:powerlawexp}
\end{figure}

We introduce the power law exponent, $\gamma$, as a new parameter to the MCMC parameter chains. 
A uniform prior is imposed from 0.1 to 1.4. 
The upper edge is chosen to prevent overlap with the $\ell^2$ scaling of the Poisson term. 
The constraints on $\gamma$ are shown in Figure \ref{fig:powerlawexp}. 
The data prefer $\gamma =0.81 \pm 0.08$, consistent with the nominal value of 0.8. 
Adding the  \planck\ 353 GHz CIB bandpowers improves the shape measurement to $\gamma = 0.82 \pm 0.05$. 
The likelihood of the best-fit point is not improved by the introduction of this new parameter for either set of data. 
Both results are consistent with the recent constraint of $\gamma =0.75 \pm 0.06$ by \citet{addison11}  using data from BLAST, \planck, and ACT. 
For simplicity and ease of comparison with earlier results, we continue to use $\gamma =0.8$ in all other parameter chains. 

\begin{figure}[tbh]\centering
\includegraphics[width=0.45\textwidth]{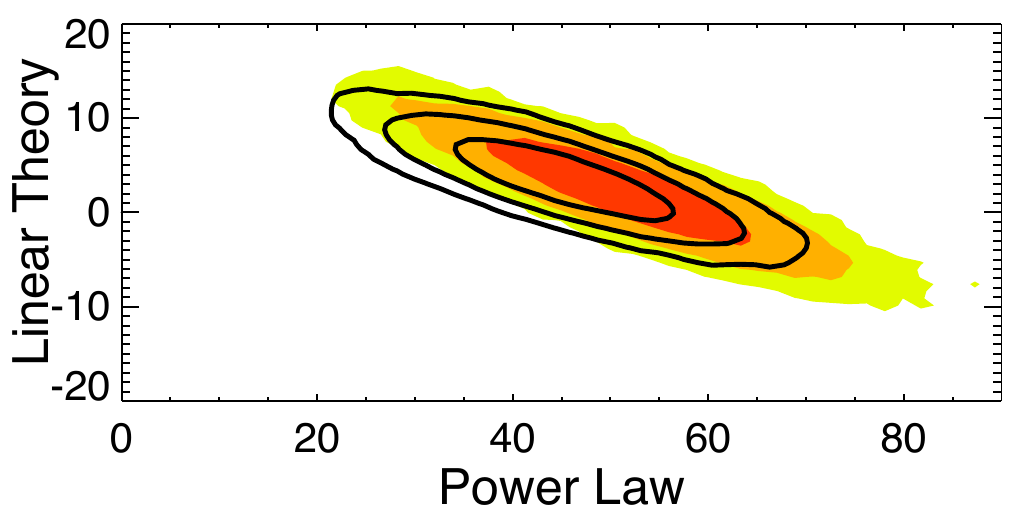}
  \caption[]{
  2D likelihood function for the amplitudes in $\mu{\rm K}^2$ at 220\,GHz of the power-law and linear-theory CIB clustering terms. 
  The {\bf filled yellow to red contours} show the 1, 2, and 3\,$\sigma$ constraints with  SPT+\planck\ data at 95 - 220\,GHz.
  The {\bf black contours} show the same constraints when including the \planck\ 353\,GHz CIB bandpowers.
  In both cases, the data significantly prefer the power law clustering template, and are consistent at 1\,$\sigma$ with zero additional linear theory power. 
   }
  \label{fig:powerlawlinear}
\end{figure}

We also test introducing the linear theory model with a free amplitude in addition to the power law term with $\gamma = 0.8$. 
The constraints on the amplitudes  of the linear theory and power law terms at 220\,GHz and $\ell = 3000$ are shown in Figure \ref{fig:powerlawlinear}. 
We do not find a significant preference for a non-zero amplitude of the linear theory term; the derived $1\,\sigma$ confidence intervals are [-0.02, $0.08\,\mu{\rm K}^2$], [-0.2, $0.7\,\mu{\rm K}^2$], and [-1.4, $6.2\,\mu{\rm K}^2$] at 95, 150, and 220\,GHz respectively. 
We have allowed negative values to allow the maximum shape flexibility; the total power in the two clustering terms is always positive. 
Note that these ranges should be compared to $ 0.67\pm 0.13$, $5.7\pm 1.0$, and $51\pm9\,\mu{\rm K}^2$ for the corresponding amplitude of the power-law term. 
The quality of the fit is not  improved by adding this free parameter. 
The preferred values imply that most clustered CIB power is in the power law shape instead of the linear theory shape at $\ell \gtrsim 500$. 
Adding the \planck\ 353 GHz bandpowers does not significantly change this picture.

\subsection{CIB constraints with tSZ-CIB correlation}

The galaxy clusters that contribute to the tSZ power spectrum have a higher than average density of galaxies, each with some amount of star formation. 
This will lead to an anti-correlation between the tSZ and CIB power spectra. 
We introduce this correlation following the treatment in \S\ref{subsec:tszcib}. 
Note that in this treatment, the correlation is defined with respect to all CIB power, although the Poisson and clustered components may correlate differently. 
As discussed in Z11, some models predict a correlation effectively independent of $\ell$ at current instrumental sensitivities  (e.g. \citealt{sehgal10}), while others predict that the magnitude of the correlation will rise across the relevant multipoles. 
We use a single correlation coefficient on all angular scales. 
The data prefer a correlation coefficient of $\xi = -0.18 \pm 0.12$ (i.e., an anti-correlation 
between tSZ and CIB signals, consistent with a fraction of the CIB emission being spatially coincident with 
the tSZ decrement from clusters). 
This constraint depends on the assumption that the correlation is independent of $\ell$. 
We tested a correlation model described by Z11 where the absolute magnitude of the correlation doubles from $\ell = 2000 - 10000$, and found similar uncertainties but a mean value $1\,\sigma$ closer to zero at $\ell = 3000$. 
In the near future, we expect observations with {\it Herschel} and {\it Planck} in combination with SPT to place better observational constraints on this correlation.

Introducing a tSZ-CIB correlation does very little to the inferred CIB powers; 
however, the uncertainties increase by 10-50\%. 
With tSZ-CIB correlations, the constraint on the Poisson power at 150\,GHz is $D^{p}_{3000}=8.04 \pm 0.48 \, \mu {\rm K}^2$, and the constraint on the clustered power at 150\,GHz is $D^{c}_{3000}=6.71 \pm 0.74 \, \mu {\rm K}^2$. 
The best-fit spectral index is lower by $1\,\sigma$, and the uncertainty increases by 50\% to give $\alpha = 3.45 \pm 0.11$. 
However as discussed earlier, a free correlation significantly degrades the kSZ constraint.

\section{Conclusions}
\label{sec:conclusion}

We have presented the CMB temperature anisotropy power spectrum from the
complete $800\,$\sqdeg\ of sky observed by SPT in the 2008 and 2009 austral winters.
The six cross-spectra from observations at three frequencies (95, 150, and $220\,$GHz) are shown in
Table~\ref{tab:bandpowers}.  
The bandpowers measure arcminute-scale ($\ell > 2000$) millimeter wavelength anisotropy. 
On these small scales, DSFGs, radio galaxies, and the kinetic and thermal SZ effects contribute significant power in addition to the primary CMB anisotropy.

Using MCMC methods, we perform multi-frequency fits to a combined data set including WMAP7, BAO, H$_0$, previously published low$-\ell$ SPT bandpowers, \planck\ CIB bandpowers, and the $\ell > 2000$ SPT bandpowers presented in this work. 
We find that the minimal model needed to explain the current data adds five free parameters beyond the six $\Lambda$CDM parameters: four describing DSFGs and one describing the amplitude of the tSZ power spectrum. 
We also include a free parameter for the amplitude of the kSZ spectrum, and a prior on a Poisson population of radio galaxies.
This six parameter extension improves the  likelihood of the best-fit model by $\delchisq = -2081$. 
We explore a number of extensions to this baseline model, including different templates for the kSZ, tSZ, or CIB terms, different modeling of the CIB frequency spectrum, and correlations between the tSZ and CIB. 
We do not find these extensions significantly improve the quality of fits to the data, however, they may slightly alter the conclusions drawn.

We fit for the amplitude of the tSZ and kSZ power spectra, using a single amplitude parameter defined in units of $\mu{\rm K}^2$ at $\ell=3000$ and $150\,$GHz. 
The constraint on the 150\,GHz tSZ power is $D^{\rm tSZ}_{3000}=3.65\pm 0.69\, \mu {\rm K}^2$ in the baseline model. 
We find the tSZ amplitude is independent of the template shape for the two models considered: the Shaw and Sehgal tSZ models.
We find the kSZ results are also robust with respect to template shape. 
The 95\% confidence upper limit on the kSZ power is $D^{\rm kSZ}_{3000}<2.8\, \mu {\rm K}^2$.

We test whether extensions to the primary CMB modeling, including running of the scalar index, changing number of neutrino species, or massive neutrinos, affect the SZ constraints and find they do not. 
We also investigate whether alternative CIB modeling assumptions lead to changes in the SZ constraints. 
The SZ constraints are fairly insensitive to the CIB modeling. 
An exception is their sensitivity to the modeling of the tSZ-CIB correlation. 
When allowing for the possibility of this correlation, the allowed parameter space expands along the degeneracy line of $D^{\rm tSZ}_{3000} + 0.5 D^{\rm kSZ}_{3000}$. 
The intersection of this line moves by less than $1\,\sigma$: $D^{\rm tSZ}_{3000} + 0.5 D^{\rm kSZ}_{3000} = 4.15 \pm 0.56\, \mu {\rm K}^2$ without correlations and $4.60 \pm 0.63\, \mu {\rm K}^2$ with free correlations. 
We note that this degeneracy line closely resembles that found by L10, who used a linear 
combination of the $150\,$GHz and 220\,GHz maps to subtract the CIB. 
When correlations are allowed,  the 95\% confidence upper limits on the total kSZ power are $D^{\rm kSZ}_{3000}<6.7\, \mu {\rm K}^2$ and $D^{\rm kSZ}_{3000}<5.7\, \mu {\rm K}^2$ for the homogeneous and patchy reionization templates respectively. 
The real kSZ signal is expected to be an admixture of the two templates.

We use the measured tSZ power to constrain $\sigma_8$ and the sum of
the neutrino masses.  For massless neutrinos and the optimistic assumption of 
assuming perfect tSZ modeling, the tSZ power constraint tightens the
CMB+BAO+H$_0$ $\sigma_8$ constraint by 40\%.  However, the preferred
$\sigma_8$ value is $3\,\sigma$ discrepant between the two (Shaw and Sehgal) tSZ models
considered.  With a conservative 50\%
modeling uncertainty, adding the tSZ information does not significantly improve the constraints on
 $\sigma_8$.  
The impact of the
tSZ power measurement is more significant when neutrinos are allowed
to have mass.  
With massive neutrinos, even with a 50\%
modeling uncertainty, the tSZ power spectrum reduces the $\sigma_8$
uncertainty by 30\% and tightens the upper limit on the total neutrino
mass from 0.52\,eV to 0.40\,eV.  Precise measurements of the pressure
profiles of high redshift clusters are essential to reduce the theory
uncertainty and realize the statistical power of the tSZ measurement.

We also use the SPT data to constrain the astrophysical foregrounds (after masking the brightest point sources). 
Two parameters -- which can be fixed with priors from other observations -- fit the radio source component. 
We find a four parameter model is adequate to describe the Poisson and clustered DSFGs. 
The spectrum of the DSFGs falls sharply from 220 to $150\,$GHz, with an effective spectral index of $\alpha = 3.56 \pm 0.07$. 
This is steeper by $\sim$$2\,\sigma$ than extrapolations from models based on higher frequency observations of the CIB.
However, it is in agreement with previous spectral index measurements at these frequencies by S11, \citet{dunkley11}, and \citet{addison11}. 
We find Poisson DSFG powers at $\ell = 3000$ of $D^{p}_{3000}=7.54 \pm 0.38\, \mu {\rm K}^2$ and $D^{p}_{3000}=68.0 \pm 3.3, \mu {\rm K}^2$ at 150 and 220\,GHz respectively.
We find clustered DSFG powers at $\ell = 3000$ of $D^{c}_{3000}=6.25 \pm 0.52\, \mu {\rm K}^2$  and $D^{c}_{3000}=56.3 \pm 4.2\, \mu {\rm K}^2$ at 150 and 220\,GHz respectively. 
We find some preference for different spectral indices for the clustering and Poisson DSFG components when we exclude the \planck\ 353\,GHz data; this preference vanishes when the 353\,GHz data is included.

With \planck\ and SPT data, the CIB  is constrained on angular scales from $\ell = 80 - 9400$ at 220\,GHz. 
This large lever arm allows both the separation of the Poisson and clustered DSFG components, and a tight constraint on the shape of the clustering component. 
We find that the data prefer the clustered power to be predominately in the form of a power law $\ell^{\gamma}$ rather than the linear theory prediction. 
This is strong evidence for non-linearities in the DSFG clustering. 
We further constrain the exponent of the power law to be $\gamma = 0.82 \pm 0.05$, consistent with other recent constraints \citep{addison11}.

We test the impact of introducing a tSZ-CIB correlation coefficient, and find the data prefer a negative correlation at 98\% confidence. 
Note that this significance may vary for more sophisticated CIB models. 
The tSZ-CIB anti-correlation modestly increases the inferred CIB powers, while counter-intuitively reducing the tSZ power and increasing the allowed kSZ power. 
The uncertainties on both SZ terms increase significantly when correlations are allowed. 
External data or modeling to constrain the tSZ-CIB correlation would be extremely valuable to future kSZ constraints. 

A companion paper, Z11, uses the SPT bandpowers and kSZ upper limits presented here to investigate the epoch of reionization. 
Z11 constrains the duration of reionization from the SPT measurement of kSZ power. 
The midpoint of reionization is determined from the WMAP constraint on the optical depth. 
Together, the measurements yield the first constraints on the beginning, duration, and end of reionization from the CMB. 

The SPT survey of 2500\,\sqdeg\ finishes in November, 2011. 
The survey area, comprising 6\% of the total sky, has been mapped to depths of approximately 40, 18, and $70\,\mu$K-arcmin at 95, 150, and $220\,$GHz respectively. 
We expect bandpower uncertainties for the full dataset to be reduced by a factor of 1.7 compared to this work. 
We expect from simulations that the constraints on the tSZ, kSZ, and tSZ-CIB correlation will improve commensurately. 
In addition, we have selected 100\,\sqdeg\ of the SPT survey for deeper integration and  \herschel\  followup. 
\herschel/SPIRE will observe this region in the winter of 2011. 
The combined SPT/\herschel\ dataset will be an invaluable resource in understanding the CIB and its correlation with the SZ power spectra. 
We will be able to use the \herschel\ maps to clean the CIB from the SPT
maps of this 100\,\sqdeg\ field, further improving measurements of the tSZ and kSZ power spectra. 
These  results will advance our understanding of structure formation and the
reionization of the universe.

\acknowledgments

We thank Cien Shang, Guilaine Lagache, Olivier Dore, and Marco Viero for useful discussions and assistance with the CIB modeling. 
The South Pole Telescope is supported by the National Science
Foundation through grants ANT-0638937 and ANT-0130612.  Partial
support is also provided by the NSF Physics Frontier Center grant
PHY-0114422 to the Kavli Institute of Cosmological Physics at the
University of Chicago, the Kavli Foundation and the Gordon and Betty
Moore Foundation.  
The McGill group acknowledges funding from the National   
Sciences and Engineering Research Council of Canada, 
Canada Research Chairs program, and 
the Canadian Institute for Advanced Research. 
R. Keisler acknowledges support from NASA Hubble Fellowship grant HF-51275.01.
B.A. Benson is supported by a KICP Fellowship.
M. Dobbs acknowledges support from an Alfred P. Sloan Research Fellowship.
L. Shaw acknowledges the support of Yale University and NSF grant AST-1009811.  
M. Millea and L. Knox acknowledge the support of NSF grant 0709498.  
This research used resources of the National Energy Research Scientific Computing Center, which is supported by the Office of Science of the U.S. Department of Energy under Contract No. DE-AC02-05CH11231. Some of the results in this paper have been derived using the HEALPix \citep{gorski05} package. We acknowledge the use of the Legacy Archive for Microwave Background Data Analysis (LAMBDA). Support for LAMBDA is provided by the NASA Office of Space Science.


\bibliography{../../BIBTEX/spt.bib}

\end{document}